\documentclass[aps,pra,preprint,groupedaddress]{revtex4-1}
\usepackage{graphicx}
\usepackage{amsmath}
\usepackage{amsfonts}
\usepackage{amssymb}
\usepackage{mathrsfs}
\usepackage{bm}
\usepackage{caption}
\usepackage{subcaption}
\usepackage{multirow}
\usepackage{placeins}

\newcommand{\eps}{\epsilon}

\newcommand{\br}{{\bf r}}
\newcommand{\bx}{{\bf x}}
\newcommand{\by}{{\bf y}}
\newcommand{\bk}{{\bf k}}

\newcommand{\bn}{{\bf n}}

\newcommand{\bp}{{\bf p}}

\newcommand{\bv}{{\bf v}}
\newcommand{\bq}{{\bf q}}
\newcommand{\bA}{{\bf A}}
\newcommand{\bB}{{\bf B}}
\newcommand{\bE}{{\bf E}}
\newcommand{\bG}{{\bf G}}
\newcommand{\bK}{{\bf K}}

\newcommand{\bR}{{\bf R}}

\DeclareMathAlphabet{\mathcalligra}{T1}{calligra}{m}{n}
\DeclareMathAlphabet{\mathpzc}{OT1}{pzc}{m}{it} 
\begin{document}

\title{Dirac Fermions in Solids --- from High Tc cuprates and Graphene to Topological Insulators and Weyl Semimetals.}


\author{Oskar Vafek}
\affiliation{National High Magnetic Field Laboratory and Department of Physics, Florida State University, Tallahassee, Florida 32306, USA}
\author{Ashvin Vishwanath}
\affiliation{Department of Physics, University of California, Berkeley, California 94720, USA}

\begin{abstract}
Understanding Dirac-like Fermions has become an imperative in modern
condensed matter sciences: all across its research frontier, from
graphene to high T$_c$ superconductors to
the topological insulators and beyond, various electronic systems
exhibit properties which can be well described by the Dirac equation.
Such physics is no longer the exclusive domain of quantum field
theories and other esoteric mathematical musings; instead, real
physics of real systems is governed by such equations, and important
materials science and practical implications hinge on our
understanding of Dirac particles in two and three dimensions. While
the physics that gives rise to the massless Dirac Fermions in each
of the above mentioned materials is different, the low energy
properties are governed by the same Dirac kinematics.
The aim of this article is to review a selected cross-section of this vast field
by highlighting the generalities, and contrasting the specifics, of several physical systems.
\end{abstract}

\maketitle

\tableofcontents

\makeatletter
\let\toc@pre\relax
\let\toc@post\relax
\makeatother


\section{Dirac, Weyl, and Majorana}
\begin{quotation}ÒI think it is a peculiarity of myself that I like to play about with equations, just looking for beautiful mathematical relations which maybe don't have any physical meaning at all. Sometimes they do.Ó  - Paul A. M. Dirac (1902 - 1984)\end{quotation}

Published in 1928 by Paul Dirac\cite{Dirac1928}, the eponymous equation is among the finest achievements of human intellect. The equation, now taught in virtually every physics department around the world, has brought together Einstein's special theory of relativity and quantum mechanics. It led to the prediction of antimatter, namely the positron as the electron's anti-partner. It casted the spin-$1/2$ nature of the electron in a new light, and is now a key building block of the Standard Model of particle physics.
For a free particle, it can be written as
\begin{equation}\label{eq:Dirac equation}
i\hbar\frac{\partial}{\partial t}\psi=\left(c\boldsymbol\alpha\cdot\bp+\beta mc^2\right)\psi,
\end{equation}
where the momentum operator $\bp=-i\hbar\nabla=(p_x,p_y,p_z)$, $m$ is the mass of the particle, $c$ is the speed of light in vacuum, and $\psi$ is a $4$-component object, a spinor.
There are many equivalent ways to write down the Dirac $4\times 4$ matrices; utilizing the outer product\cite{WilczekNatPhys2009} of the Pauli matrices\footnote{As usual, $\tau_1=\sigma_1=\left(\begin{array}{cc}0 & 1\\ 1 & 0\end{array}\right)$, $\tau_2=\sigma_2=\left(\begin{array}{cc}0 & -i\\ i & 0\end{array}\right)$, and $\tau_3=\sigma_3=\left(\begin{array}{cc}1 & 0\\ 0 & -1\end{array}\right)$.}, one such way is
$\boldsymbol\alpha=\left(\tau_3\otimes\sigma_1,\tau_3\otimes\sigma_2,\tau_3\otimes\sigma_3\right)$, and $\beta=-\tau_1\otimes 1$.
The equation was originally intended for the electron, which is, of course, a massive, spin-$1/2$, charged particle, i.e., a Dirac Fermion.

There is a certain degree of simplification occurring in this equation in the special case of massless particles.
All three $\boldsymbol\alpha$ matrices are block diagonal, while the term proportional to the mass is block off-diagonal.
Therefore, if we consider massless particles, the right-hand-side of the Dirac equation no longer couples the upper two components of $\psi$,
let's call them $\chi_+$, and the lower two components, $\chi_-$. Thus, with $m=0$, it can be written in a simpler form
\begin{eqnarray}\label{eq:Weyl equation}
i\hbar\frac{\partial}{\partial t}\chi_{\pm}=\pm c\boldsymbol\sigma\cdot \bp \chi_{\pm}.
\end{eqnarray}
This is the Weyl equation\cite{Weyl1929} and $\chi$'s are referred to as Weyl Fermions.

Both of these equations involve real and complex numbers.
Majorana noticed\cite{Majorana1937} that it is possible to write the Dirac equation --- including the mass term --- entirely in terms of real numbers\cite{WilczekNatPhys2009}.
This can be accomplished by choosing the $\boldsymbol\alpha$ matrices to be purely real and the $\beta$ matrix to be purely imaginary,
because then both the right-hand-side and the left-hand-side of the Dirac equation are purely imaginary. For example, $\boldsymbol\alpha=\left(-\tau_1\otimes\sigma_1,\tau_3\otimes 1,-\tau_1\otimes\sigma_3\right)$, and $\beta=\tau_1\otimes \sigma_2$ does the job. Once the equation is purely real, its solutions can also be chosen to be purely real.
In quantum field theory, a real field describes a particle which is its own antiparticle.

This review is about how such equations provide an accurate description of some 2- and 3-dimensional non-relativistic systems, where Dirac or Weyl Fermions emerge as low energy excitations.
It is also about how these excitations behave when subjected to external fields,
and how to relate the perturbing ``potentials'' (e.g. scalar, vector, mass etc.) appearing in the effective Dirac equation to either externally applied fields produced in a laboratory, or to defects and impurity potentials. A few consequences of many-body interactions will also be reviewed.
We will not discuss any of the fascinating aspects of Majorana Fermions in condensed matter; this topic has already been covered in Ref.\cite{Beenakker2013} and references therein.
The main topics of this paper form a vast area of physics, and we ask the reader to keep in mind that it is impossible to do it justice in the review with a given allotted space.

\section{When and why to expect Dirac points in condensed matter?}

In a non-relativistic condensed matter setting, the time evolution of any many body state $|\Psi\rangle$ is governed by the Schrodinger equation
\begin{eqnarray}
i\hbar \frac{\partial}{\partial t}|\Psi\rangle=\mathcal{H}|\Psi\rangle
\end{eqnarray}
where $\mathcal{H}$ is the Hamiltonian operator. This Hamiltonian contains the kinetic energy of the electrons and ions, as well as any interaction energy among them.
Our aim is to illustrate how and when we may expect the relativistic-like Dirac dispersion to arise from $\mathcal{H}$ in a cold non-relativistic solid state. We do so first by pure symmetry considerations and then in a brief survey of several physical systems realizing Dirac-like physics.
We will assume that the heavy ions have crystallized and to the first approximation let us ignore their motion. As such, their role is solely to provide a static periodic potential which scatters the electron Schrodinger waves and, if the spin-orbit coupling is also taken into account, the electron spins.
Then $\mathcal{H}\rightarrow \mathcal{H}_{0}+\mathcal{H}_{int}$, where $\mathcal{H}_0$ includes all one body effects and $\mathcal{H}_{int}$ all many-body electron-electron interaction effects.

According to the Bloch theorem, the energy spectrum $E_{n}({\bf k})$ and the eigenstates $|\phi_{n,\bk}\rangle$ of $\mathcal{H}_0$ can be described by a discrete band index $n$ as well as a continuous $D$-dimensional vector ${\bf k}$, the crystalline momentum, which is defined within the first Brillouin zone.
Consider now two distinct but adjacent energy bands $E_{n+}({\bf k})$ and $E_{n-}({\bf k})$, and assume that for some range of ${\bf k}$ the two bands approach each other, i.e. the energy difference $|E_{n+}({\bf k})-E_{n-}({\bf k})|$ is much smaller than the separation to any one of the rest of the energy bands.
One way to derive the effective Hamiltonian for the two bands is to start with a pair of (orthonormal) variational Bloch states, $|u_{{\bf k}}\rangle$ and $|v_{{\bf k}}\rangle$, consistent with, and adapted to, the symmetries of $\mathcal{H}_0$.
Then the effective Hamiltonian takes the form
\begin{eqnarray}
\mathcal{H}_{eff}=\sum_{{\bf k}}\psi^{\dagger}_{{\bf k}}H({\bf k})\psi_{\bf k}
\end{eqnarray}
where the first component of the creation operator $\psi^{\dagger}_{\bk}$ adds a particle (to the $N$-body state) in the single particle state $|u_{\bk}\rangle$ and antisymmetrizes the resulting $N+1$-body state. Similarly, the second component creates a particle in the state $|v_{\bk}\rangle$ and
\begin{eqnarray}\label{eq:Heff 2by2}
H(\bk)=\left(\begin{array}{cc}\langle u_{\bk}|\mathcal{H}_0|u_{\bk}\rangle & \langle u_{\bk}|\mathcal{H}_0|v_{\bk}\rangle\\
\langle v_{\bk}|\mathcal{H}_0|u_{\bk}\rangle & \langle v_{\bk}|\mathcal{H}_0|v_{\bk}\rangle
\end{array}\right) \equiv f(\bk)1_2 + \sum_{j=1}^3g_{j}(\bk)\sigma_j
\end{eqnarray}
where $1_2$ is a unit matrix and $\sigma_j$ are the Pauli matrices.
The corresponding one particle spectrum is
\begin{equation}
E_{\pm}=f(\bk)\pm\sqrt{\sum_{j=1}^3g^2_{j}(\bk)}.
\end{equation}
For a general $\bk$-point and in the absence of any other symmetries, $g_{j}(\bk)\neq 0$ for each $j$. It is clear from the expression for $E_{\pm}(\bk)$ that the two bands touch only if $g_{j}(\bk_0)=0$ for each $j$ at some $\bk_0$.

In 3D, we can vary each of the three components of $\bk$ and try to find simultaneous zeros of each of the three components of $g_j(\bk)$. To see that this may be possible without fine-tuning, note that in general each one of the three equations $g_j(\bk)=0$ describes a 2D surface in $\bk$-space. The first two surfaces may generally meet along lines, and such lines may then intersect the third surface at points without additional fine-tuning. If such points exist, they generally come in pairs and the dispersion near each may be linearized.
The effective Hamiltonian near one such point $\bk_0$ takes the form
\begin{eqnarray}
H(\bk)=E_{\bk_0}+ \hbar\bv_0\cdot(\bk-\bk_0)1_2 + \sum_{j=1}^3\hbar\bv_{j}\cdot(\bk-\bk_0)\sigma_j.
\end{eqnarray}
If $\bv_0=0$ and the three velocity vectors $\bv_j$ are mutually orthogonal this has the form of an anisotropic Weyl Hamiltonian.
Of course, far away from $\bk_0$ both bands may disperse upwards or downwards, in which case even if the Fermi level could be set to $E(\bk_0)$, there would be additional Fermi surface(s).

In 2D, only two components of $\bk$ can be freely varied, and therefore it is impossible to find simultaneous zeros of three functions $g_{j}(\bk)$ without additional fine-tuning. Simply stated, in general, three curves do not intersect at the same point. Therefore, in the absence of additional symmetries that may constrain the number of independent $g_{j}(\bk)$'s, the two levels will avoid each other.

\subsection{Dirac points and Kramer's pairs}
We have intentionally refrained from any discussion of the electron spin degeneracy, or time reversal symmetry, which were not assumed to be present in the above discussion. For a number of physical systems considered later on, the product of the time reversal and the space inversion leaves the crystalline Hamiltonian invariant.
This symmetry implies that, at each $\bk$, every electronic level is doubly degenerate, because if $\phi_{\bk}(\br)$ is an eigenstate, then so is its orthogonal Kramers partner, $i\sigma_2\phi^*_{\bk}(-\br)$, where $\sigma_2$ acts on the spin part of the wavefunction.
Therefore, the appropriate variational quadruplet of mutually orthogonal states describing two nearby bands can be constructed from $u_{1\bk}(\br)|\uparrow\rangle + u_{2\bk}(\br)|\downarrow\rangle$, its Kramers partner $-u^*_{1\bk}(-\br)|\downarrow\rangle+u^*_{2\bk}(-\br)|\uparrow\rangle$, and $v_{1\bk}(\br)|\uparrow\rangle + v_{2\bk}(\br)|\downarrow\rangle$, with its partner $-v^*_{1\bk}(-\br)|\downarrow\rangle+v^*_{2\bk}(-\br)|\uparrow\rangle$.
In this four-dimensional subspace
\begin{eqnarray}\label{eq:Heff Kramer's}
H(\bk)=f(\bk)1_4+\sum_{j=1}^5g_{j}(\bk)\Gamma_j
\end{eqnarray}
where $\Gamma_1=\tau_3\otimes1$, $\Gamma_2=\tau_1\otimes1$, $\Gamma_3=\tau_2\otimes\sigma_3$, $\Gamma_4=\tau_2\otimes\sigma_1$, and $\Gamma_5=\tau_2\otimes\sigma_2$; the first Pauli matrix acts within the $u$,$v$ space and the second within the Kramers doublets.
While the corresponding one particle spectrum,
$E_{\pm}=f(\bk)\pm\sqrt{\sum_{j=1}^5g^2_{j}(\bk)}
$, exhibits a two-fold degeneracy at any $\bk$, an intersection of two Kramers pairs requires finding simultaneous zeros of five $g_j(\bk)$'s.
Clearly, the bands avoid each other because, even in 3D, this condition cannot be satisfied without additional symmetry. For example, if the spin-orbit interaction can be neglected and time reversal symmetry is preserved --- based on our earlier assumptions, this also implies that space inversion is preserved --- then the spin $SU(2)$ symmetry forces $g_3=g_4=g_5=0$. With such additional symmetry, in 3D, the accidental degeneracy may happen along 1D $\bk$-space curves and in 2D, at nodal points.

\subsection{Fermion doubling: Nielsen-Nynomiya theorem and ways around it}
The Nielsen-Nynomiya theorem states that it is impossible to construct a non-interacting lattice hopping model with a net imbalance in the number of (massless) Dirac Fermions with positive
and negative chirality, provided that certain weak restrictions apply. For example, the translationally invariant hopping amplitudes are assumed to decay sufficiently fast so that in momentum space the Hamiltonian is continuous. The full proof\cite{NielsenNynomiya1981} makes use of homotopy theory and is beyond the scope of this review; pedagogical discussion of this ``no-go'' theorem can be found in \cite{ItzyksonDrouffe1989}.
Here we will illustrate the basic idea behind it in a simple example in two space dimensions.

\begin{figure}[t]
\includegraphics[width=0.5\textwidth]{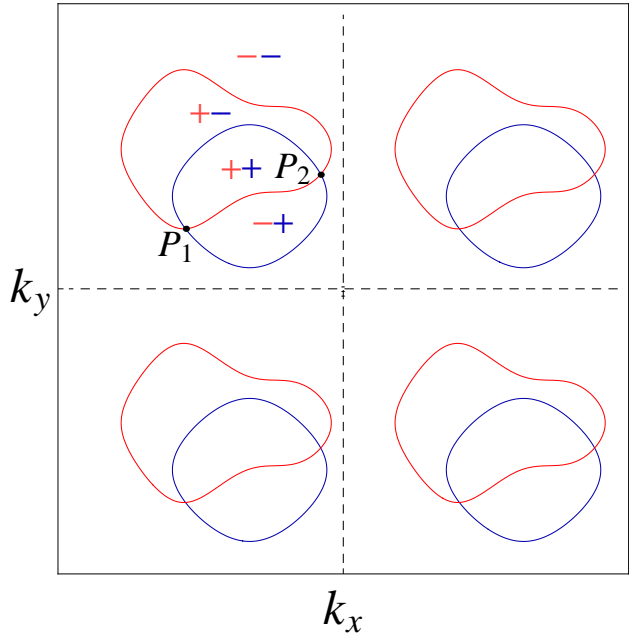}
\caption{Illustration of the Fermion doubling in the 2D lattice hamiltonian. The blue and red lines correspond to the solutions of $g_{1}(\bk)=0$ and $g_{2}(\bk)=0$, respectively. Both $g_1(\bk)$ and $g_{2}(\bk)$ are smooth and must be periodic (for illustration only 4 Brillouin zones are shown). Note that there is always an even number of intersections unless the two curves just touch. If we think of the two signs as points in the complex plane, we see that the gapless points have opposite chirality. Imagine displacing, say, the blue curve down, holding the red curve fixed. The two points $P_1$ and $P_2$ will move towards each other, and meet when the two curves touch. In this case, one of the Dirac velocities vanishes and we do not have a Dirac Fermion at all. Therefore, in any lattice formulation with finite range hopping, there will always be an even number of --- in general anisotropic --- massless Dirac Fermions with opposite chirality.}
\label{fig:Fermion doubling}
\end{figure}

Consider a model with two bands which may touch, such as the one given in Eq.(\ref{eq:Heff 2by2}) with $g_3(\bk)=0$. Then, $g_1(\bk)$ and $g_2(\bk)$ are smooth periodic functions of $k_x$ and $k_y$. If the first function vanishes along some curve in the Brillouin zone, say the one marked by red in Fig.\ref{fig:Fermion doubling}, and the second vanishes along another curve, blue in Fig.\ref{fig:Fermion doubling}, then the places where the two curves intersect correspond to massless Dirac Fermions. Periodicity guarantees that any intersection must occur at an even number of points, corresponding to an even number of massless Fermions; just touching the two curves does not produce a Dirac Fermion because at least one component of the velocity vanishes. Importantly, there is an equal number of partners with opposite chirality.

One way to remove half of the massless Fermions is to bring back $g_3(\bk)$ and to force it to vanish at only half of the intersections of the red and the blue curves in Fig.\ref{fig:Fermion doubling}. This gaps out the unwanted Dirac points, leaving an odd number of gapless points. Haldane's model for a quantum Hall effect without Landau levels is a condensed matter example where such an effect occurs along the phase boundaries separating quantum Hall phases and trivial insulating phases\cite{Haldane1988}. HgTe quantum wells are another example\cite{ButtnerNatPhys2011}; there such ``single valley'' massless Dirac Fermions have been experimentally realized at the phase boundary separating the quantum spin Hall phase\cite{Maciejko2011} and a trivial insulating phase. In the lattice regularization of the relativistic high energy theory, for which the space-time points are discrete and separated by at least a lattice constant $a$, a similar term corresponds to the so-called Wilson mass term: a 4-momentum dependent mass, $\sum_{j=0}^4\Delta\left(1-\cos (k_ja)\right))$, which vanishes at $\bk=0$ and $\omega=0$.
Adding the Wilson mass results in only one massless Fermion, but it is not chiral. Moreover, in any condensed matter setting, making the $\bk$-dependent mass term vanish at an isolated $\bk$-point requires fine tuning, and therefore such gapless points generally correspond to phase boundaries as opposed to phases\cite{Haldane1988}\cite{Maciejko2011}.

\subsubsection{Domain wall Fermions and 3D topological insulators}
Another way of avoiding the Fermion doubling on the lattice has been well known in high energy theory\cite{Kaplan92}\cite{Kaplan2012}.
Kaplan's idea has been to start with massive Fermions and to make a mass domain wall along the non-physical $4$th spatial dimension, hereby labeled by $w$. By mass domain wall we mean that for positive $w$ the mass is $m_0$, and for negative $w$ it is $-m_0$. For the $w=0$ lattice site the mass vanishes. To this domain wall mass term add a Wilson mass term. There is then a range of values of $m_0$ for which we have a single chiral 3+1D massless Dirac, i.e. Weyl, particle on the domain wall. For $m_0<2\Delta$ this can be understood as the two sides having a mass inversion at only one $\bk$-point, namely at the origin. This was proposed as a method to simulate --- on a lattice --- chiral Fermions in odd space-time dimensions: from 4+1D to 3+1D or from 2+1D to 1+1D.

Unlike the Wilson mass, its condensed matter reincarnation is frequency independent, although of course momentum dependent.
Massless domain wall Fermions have been discussed by Volkov and Pankratov at a 2D interface between (3D) SnTe and PbTe\cite{VolkovPankratov1985}.
Such massless Dirac Fermions are similar to those appearing at the surface of strong 3D topological insulators, although there is a difference:
in the former case the mass sign change occurs at an even number of points in the Brillouin zone while in the latter at an odd number\cite{FuKaneMele2007}\cite{FuKanePRB2007}.

\section{Dirac particles subject to external perturbations}
\label{sec:external perturbations}

For relativistic Dirac Fermions described by $4$-component spinors, external perturbations take the form of space-time dependent $4\times4$ matrices, which we denote by $V(\br,t)$. In the Hamiltonian formalism
\begin{eqnarray}
H=\int d^3\br \psi^{\dagger}(\br)\left(c\boldsymbol\alpha\cdot\bp+mc^2\beta+V(\br,t)\right)\psi(\br).
\end{eqnarray}
There are 16 linearly independent $4\times4$ matrices which can be chosen for $V(\br,t)$. In a relativistic context, their physical meaning is determined by their properties under Lorentz transformations.
\begin{enumerate}
   \item If the matrix structure of $V(\br,t)$ is the same as $\beta$, it clearly acts as a space-time varying mass; because it is a scalar under the Lorentz transformation it is also sometimes referred to as a scalar potential\cite{Thaller1992}.
   \item Any $V(\br,t)$ of the form $-e\boldsymbol\alpha\cdot \bA(\br,t)$ acts as the spatial component of the electro-magnetic vector potential; it enters via minimal coupling.
   \item If $V(\br,t)=e\Phi(\br,t)$, then it corresponds to the time component of the electro-magnetic potential, or electrical potential.
   \item Of the 11 remaining matrices, 6 are Lorentz tensor fields, 4 are pseudo-vectors and 1 is pseudo-scalar\cite{Thaller1992}.
\end{enumerate}
Before proceeding, it is important to stress that the appropriate $V$ --- which describes how Dirac Fermions in a given condensed matter system react to, say, an external {\it physical} magnetic field --- depends on the system itself. For example, it is not the same in graphene and $d$-wave superconductors. This will be elaborated on in later sections.

As mentioned earlier, for massless Dirac Fermions the kinetic energy term, $\boldsymbol\alpha\cdot\bp$, can be chosen to be block diagonal. If the external perturbation $V(\br,t)$ does not couple the two Dirac points, then such perturbation is also block diagonal. In 2D --- where $\bp$ is a 2-component vector --- within each $2\times2$ block such perturbation can be identified to be either a mass, or a 3-component electro-magnetic potential, $\bA=(\Phi,A_x,A_y)$. A constant mass term opens a gap in the spectrum; this gap may close at the boundaries or defects, but persists in their absence. Simply put, for any energy $-m<E<m$, the equation $E^2=c^2\bp^2+m^2$ forces $\bp$ to be imaginary and the corresponding states can at best be evanescent. A constant electric potential, $\Phi$, shifts the energy eigenvalues; the constant space components, $A_{x}$ or $A_y$, shift the momentum. The situation is similar in 3D, except the $2\times 2$ matrix, which in 2D could be identified with the mass-like term, does not open a gap in 3D. Rather, it also shifts the momentum, and therefore should be thought of as another space component of the the vector potential.

Such simple intuitive arguments\cite{LudwigPRB1994} show why Dirac particles can be confined by a spatially varying mass, but not by a spatially varying electric potential. This observation is behind the famous Klein ``paradox''\cite{YoungKim2011}. Instead of confining the massless Dirac particles, such an electric potential causes a transfer of states towards the Dirac point, a situation loosely analogous to an impurity electric potential creating midgap states in semiconductors.

\begin{figure}[t]
\centering
\includegraphics[height=8cm]{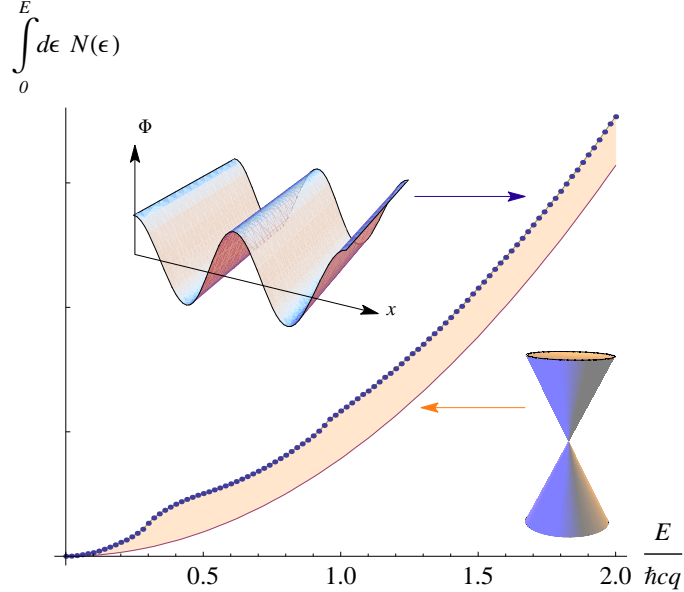}
\caption{Integrated single particle density of states for a massless Dirac Fermion in 2D subject
to a static 1D periodic electric potential $\Phi_0\cos\left(qx\right)$, blue dots, where $\Phi_0=\hbar cq$; solid line is for a free massless Dirac particle.
Note the buildup of the spectral weight which is recovered only near the cutoff energy, much larger than the scale shown.}
\label{fig:Diracelectricpotential}
\end{figure}

A uniform electric field, $\bE=-\nabla\Phi$, accelerates charged massless Dirac particles and leads to non-equilibrium phenomena; it produces charge electron-positron pairs out of the filled Dirac sea via the Schwinger mechanism\cite{SchwingerPR1951}. For massless Dirac particles in 2D such rate has been calculated to be $\sim (eE)^{3/2}$ \cite{SchwingerPR1951}\cite{AllorCohenMcGady} and, argued to lead to electrical current increasing as $E^{3/2}$ above a finite field scale below which it is $E$-linear\cite{DoraMoessnerPRB2010}\cite{RosensteinPRB2010}\cite{GavrilovPRD2012}.

The effect of a static 1D plane-wave electrical potential, $\Phi(x,y)=\Phi_0\cos(qx)$, on 2D massless Dirac Fermions was considered in Ref. \cite{BreyFertig2009}. Based on our discussion, we intuitively expect that such potential locally shifts the Fermi energy away from the Dirac point and introduces electron-positron ``stripe puddles''. The energy spectrum has a particle-hole symmetry: for every eigenstate $\psi_E(x,y)$ with an energy $E$, there is an eigenstate $\sigma_3\psi_E\left(x+\pi/q,y\right)$ with an energy $-E$. For this result we assumed that the kinetic energy term is $c\left(p_x\sigma_1+p_y\sigma_2\right)$. The full quantum mechanical solution of this problem, performed numerically using a large number of plane-wave states, shows that, while the energy spectrum remains gapless, the spectral weight is indeed shifted towards the Dirac point. This is shown in Figure \ref{fig:Diracelectricpotential}, where we compare the integrated density of states, starting from $E=0$, in the presence and absence of the periodic potential. Clearly there is an excess number of states at low energy. Interestingly, the ``lost'' states are recovered at energies comparable to the cutoff, which is much larger than $\Phi_0$. Analogous buildup of low energy density of states underpins the interpretation of the measured low temperature specific heat of type-II nodal d-wave superconductors in an external magnetic field, discussed in a later section.

\begin{figure}[t]
\includegraphics[height=8cm]{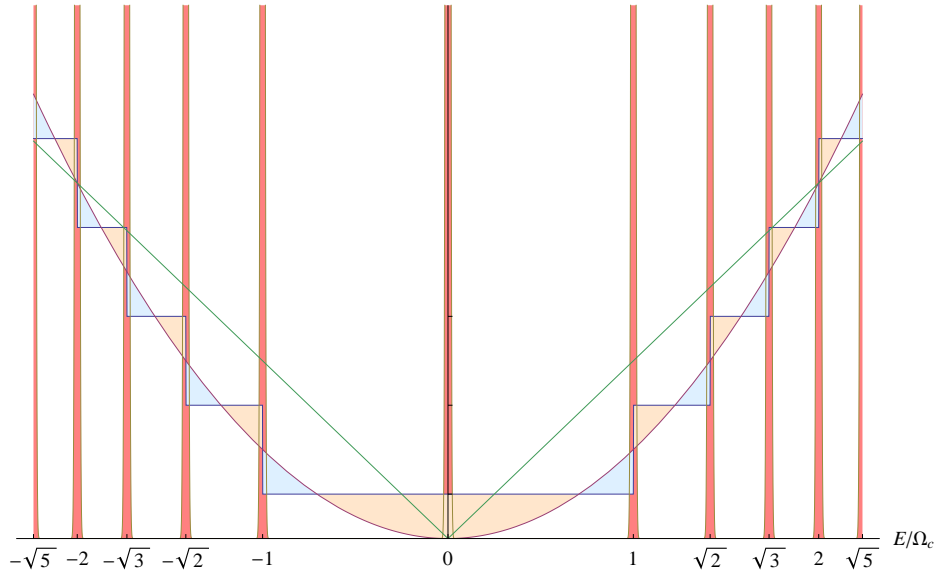}
\caption{Single particle density of states (orange) for a 2D charged massless Dirac Fermion subject to a uniform magnetic field, the Landau levels have been broadened for easier visualization; green line is the density of states for the free Dirac particle. The (step-like) integrated density of states shows that the spectral weight is redistributed over the energy window given by $\left(\sqrt{n+1}-\sqrt{n}\right)\Omega_c$ where $\Omega_c\equiv \sqrt{2}\hbar c/\ell_B$, where $\ell_B=\sqrt{\hbar c/eB}$ is the magnetic length.}
\label{fig:Diracmagneticfield}
\end{figure}

On the other hand, a uniform magnetic field directed perpendicular to the 2D plane, $B=\partial A_y/\partial x-\partial A_x/\partial y$, quantizes the electron orbits. The resulting spectrum consists of discrete Landau levels at energies $E_n=\mbox{sgn(n)}\sqrt{|n|}\Omega_c$ where $n=0,\pm1,\pm2,\ldots$, $\Omega_c=\sqrt{2}\hbar c/\ell_B$, and the magnetic length $\ell_B=\sqrt{\hbar c/eB}$; this result is easily obtained by elementary methods, see for instance \cite{CastroNetoRMP2009}. Therefore, unlike for a Schrodinger electron, the energy difference between the Landau levels of a massless 2D Dirac electron decreases with increasing energy.
Each Landau level is $N$-fold degenerate, where $N=Area/\left(2\pi\ell^2_B\right)$; the degeneracy, being proportional to the sample area, is macroscopically large.
As shown in Figure \ref{fig:Diracmagneticfield}, the uniform magnetic field causes redistribution of spectral weight over the energy interval $\left(\sqrt{n+1}-\sqrt{n}\right)\Omega_c$; the number of states which are `moved' to the Landau levels equals to the total number of states which would be present between the Landau levels in the absence of the external B-field.

\begin{figure}[t]
\includegraphics[height=8cm]{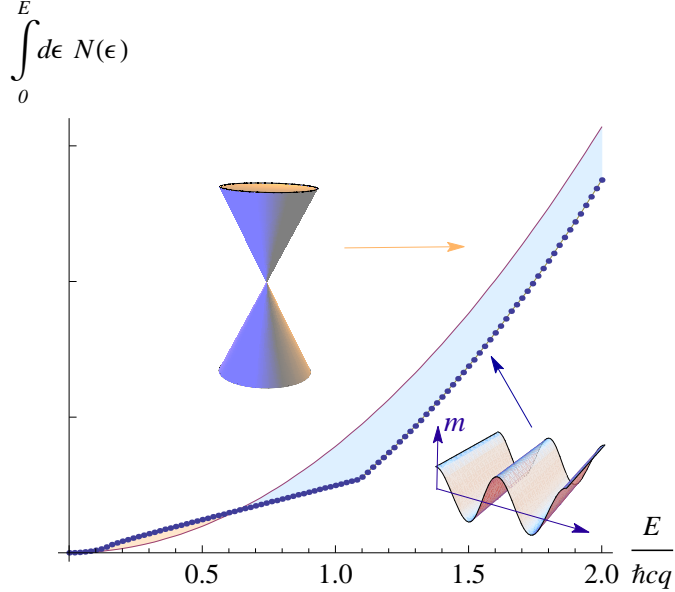}
\caption{Massless Dirac Fermion in 2D subject to the 1D periodic mass, $m(x,y)=m_0\cos(qx)$ with $m_0=cq$. Note the suppression of the spectral weight, which is recovered only near the cutoff energy, again, much larger than the scale shown.}
\label{fig:Diracmass}
\end{figure}

The effects of a perpendicular magnetic field and an in-plane electric field have been studied in the context of proving the absence of the relativistic correction to quantum Hall effect in ordinary 2D electron gas\cite{MacDonaldPRB1983}. The eigenfunctions and eigenvalues can be determined analytically, either directly\cite{MacDonaldPRB1983}, or, if $B>E$, by first Lorentz boosting the space-time coordinates and the Dirac spinors into a frame in which the electric field effectively disappears and only the Lorentz contracted magnetic field enters\cite{LukosePRL2007} [we discussed this simpler problem above] and then `inverse' Lorentz boosting the wavefunctions and eigenenergies.

Effects of non-uniform Dirac mass are quite fascinating, particularly when the mass profile is topologically non-trivial and can lead to
fractionalization of Fermion's quantum numbers. We will illustrate the effect for 1D Dirac particles, first published in 1976 by Roman Jackiw and Claudio Rebbi\cite{JackiwRebbiPRD1976}. The kinetic energy and the mass term together give $H_{JR}=c\sigma_1 p+\sigma_3 m(x)$, where $m(x)$ is fixed to approach $\pm m_0$ as $x\rightarrow\pm\infty$, vanishing once somewhere in between. One such kink configuration is, for example, $m(x)=m_0\tanh\left(x/\xi\right)$.
The spectrum of $H_{JR}$ is particle-hole symmetric, because for any state $\psi_{E}(x)$ with energy $E$, there is a state $\sigma_2\psi_{E}(x)$ with energy $-E$. As we argued earlier, any ``midgap'' state with $-m_0<E<m_0$ must be localized. Let us therefore seek states at $E=0$; they must satisfy
$i\hbar c\sigma_1 \psi'_0(x)=m(x)\sigma_3\psi_0(x)$. If we write $\psi_0(x)=\sigma_1\chi_0(x)$ and substitute, then we find $\hbar c \chi'_0(x)=m(x)\sigma_2\chi_0(x)$. The solution now follows immediately: $\chi_0(x)=N \exp\left[\frac{1}{\hbar c}\int_0^xdx'm(x')\sigma_2\right]\chi_0(0)$.
Since any $\chi_0(0)$ can be decomposed into a linear combination of the $+1$ and $-1$ eigenvectors of $\sigma_2$, we see that because the term in the integral is positive, $\chi_0$ must be purely the $-1$ eigenvector, $\left(\begin{array}{c}1\\-i\end{array}\right)$, otherwise the solution is not normalizable. There is therefore a single isolated energy level at $E=0$. For a general single kink mass profile, there may be other mid-gap states, but they must come in pairs at non-zero energies $\pm E$.

The remarkable consequence of this isolation is that if the $E=0$ midgap state is empty, while all the negative energy states are occupied with charge $e$ Fermions, then the resulting state carries an excess localized charge of $-e/2$ relative to the ground state with uniform mass without a kink. Similarly, if it is occupied, the excess charge is $e/2$. This follows from the fact that a symmetric configuration of a widely separated kink and an anti-kink leads to a pair of essentially zero energy states. In effect, one level has been ``drawn'' from the ``conduction band'' and one from the ``valence band'', each of which are missing one state. If the zero energy doublet is unoccupied, then the total charge of this state differs from the constant mass state by $-e$. Because the two localized states at the kink and the anti-kink are perfectly symmetric, we must find that the total amount of charge in the vicinity of each kink is the same, namely, $-e/2$ more than in the undistorted vacuum. If the vacuum is neutral, then each kink carries half-integral charge.
Since in any physical set up with periodic boundary conditions every kink must have a corresponding anti-kink, the quantum number fractionalization happens only locally. Globally, the charge changes by integral units.
Interestingly, if the particle-hole symmetry is weakly broken by adding to $H_{JR}$ a small constant term proportional to $\sigma_2$, then the localized states carry irrational charge\cite{RiceMele1982}.
Such ideas have fascinating applications to the physics of conducting polymers\cite{SuSchriefferHeeger1979}\cite{JackiwSchrieffer1981} and there is an extensive literature on the subject reviewed in Ref.\cite{HeegerRMP1988}.

In higher dimensions, the topologically non-trivial configurations also lead to zero modes\cite{JackiwRebbiPRD1976}\cite{JackiwRossi1981}. Just as in 1D, such results are insensitive to the details of the mass configuration, only the overall topology matters\cite{WeinbergPRD1981}.

As an illustration of an effect a non-topological configuration of the mass has on a 2D massless Dirac Fermion, we consider a 1D plane wave $m(x,y)=m_0\cos(qx)$. The resulting Hamiltonian, $c\left(p_x\sigma_1+p_y\sigma_2\right)+m(x,y)\sigma_3$, has a particle hole symmetry, in that for every eigenfunction $\psi_{E}(x,y)$ with energy $E$, there is an eigenfunction $\sigma_3\psi_{E}(x+\pi/q,y)$ with energy $-E$. The momentum along the $y$-axis, $k_y$, is conserved due to the translational symmetry in the $y$-direction. The momentum in the $x$-direction, $k_x$, is conserved only modulo the reciprocal lattice vector. At $k_x=k_y=0$ we can construct the $E=0$ state explicitly, just as we did for the Jackiw-Rebbi problem, but now both choices for $\chi_0$ lead to Bloch normalizable wavefunctions. There is therefore a doublet of states at $\bk=0$ and $E=0$. Away from $\bk=0$, there is a new anisotropic Dirac cone, with renormalized velocities. Interestingly, at $\bk=0$, the spectrum consists only of doublets at any energy because for every $\psi_{E}(x,y)$ there is $\sigma_2\psi^*_{E}(x+\pi/q,y)$ which is also at $\bk=0$, has the same energy, and is orthogonal to $\psi_{E}(x,y)$.
The overall effect on the integrated density of states is shown in Figure \ref{fig:Diracmass} for $m_0=\hbar cq$. The minimum of the $2$nd band is at $E\approx 1.1\hbar cq$ and is responsible for the change of slope. Overall, there is a suppression of the number of states at low energy --- an opposite effect compared to the electric potential case. Similarly, the ``lost'' states are recovered only at energies comparable to the cutoff, which is much larger than $m_0$.

To conclude this section, we briefly mention the chiral anomaly associated with the massless Dirac equation\cite{BellJackiw1969}\cite{Adler1969}. The anomalies in quantum field theory are a rich subject\cite{Bertlman1996} and play a very important role in elementary particle physics\cite{PeskinSchroeder1995}.
In order to illustrate the effect, note that the massless Dirac Hamiltonian in 3D and in the presence of an arbitrary external electro-magnetic field, $\int d^3\br\psi^{\dagger}(\br)\left(c\boldsymbol\alpha\cdot\left(\bp-\frac{e}{c}\bA(\br,t)\right)+e\Phi(\br,t)\right)\psi(\br)$, formally commutes with {\it both} the total particle number operator --- or equivalently, the total charge operator ---  $\int d^3\br \psi^{\dagger}(\br)\psi(\br)$, and
the total ``chiral'' charge operator $\int d^3\br \psi^{\dagger}(\br)\tau_3\otimes 1\psi(\br)$. Here we used the representation for $\boldsymbol\alpha$ used in Eq.(\ref{eq:Dirac equation}). The equation of motion for an operator $\mathcal{O}(t)$ in the Heisenberg picture is $d\mathcal{O}(t)/dt=\left[\mathcal{O}(t),H_H(t)\right]/i\hbar$, where $H_H(t)$ is the Dirac Hamiltonian in the Heisenberg representation.
Because the commutator vanishes for both the total charge and the total ``chiral'' charge, they should both be constants of motion. However, closer inspection reveals that in explicit calculations\cite{BellJackiw1969}\cite{Adler1969}\cite{PeskinSchroeder1995} an ultra-violet regularization must be adopted in order to obtain finite results. What's more, if the regularization is chosen in such a way as to maintain the conservation of charge --- a physically desirable consequence of a useful theory --- then for some configurations of electromagnetic fields, the chiral charge is {\it not} conserved and changes in time.
As an illustration, one such configuration consists of a uniform magnetic field along the $z$-direction and a parallel weak electric field\cite{PeskinSchroeder1995}. This can be described by $\Phi=0$ and $\bA(t)=\left(-By,0,A_z(t)\right)$ where the electric field is given by $-\frac{1}{c}\frac{d}{dt}A_z(t)$; the time variation of $A_z(t)$ is therefore slow. For a system with size $L^3$ and periodic boundary conditions, the momentum is quantized in units of $2\pi/L$ and the separation between the adjacent energy levels is non-zero. If the rate of change of $A_z(t)$ is much smaller than the separation of the energy levels, then we can use the adiabatic theorem, solve for the eigen-energies using the instantaneous $A_z(t)$, and then monitor the energy spectrum in time. Such an energy spectrum is easily constructed once we notice that we are effectively dealing with $\pm \sigma\cdot \left(c\bp-e\bA\right)$. These are just two copies --- with opposite sign of the Hamiltonian --- of the Landau level problem of a {\it massive} Dirac particle in 2D, with the mass set by $c\hbar k_z-eA_z(t)$. The spectrum for each is given by
$\pm \sqrt{\left(c\hbar k_z-eA_z(t)\right)^2+n\Omega^2_c}$, where $n=1,2,3,\ldots$, together with the two anomalous levels, one for each chirality, at $\pm \left(c\hbar k_z-eA_z(t)\right)$.
If, at $t=0$, we start with the many-body state where all negative energy single-particle states are occupied and all positive energy ones are empty, and then adiabatically increase $A_z$ from $0$ to $hc/eL$, then, while their {\it energy} is changing, none of the anomalous single-particle states change because their phase is locked by the periodic boundary condition. Once $A_z$ reaches $hc/eL$, we can perform the gauge transformation that removes $A_z$ from the Hamiltonian and that is consistent with the periodic boundary conditions, and find that we end up with the many-body state which appears to differ from the initial many-body state by the occupation of one additional negative chirality anomalous Landau level at energy $hc/L$ and one fewer positive chirality Landau level at energy $-hc/L$. Note that the infinitely deep negative energy Dirac sea plays a key role in this argument. Since the degeneracy of each Landau level is $L^2/2\pi\ell^2_B$, we change the difference in the number of the positive and negative chirality states, $\delta N_+-\delta N_-$, by $-2\left(L^2/2\pi\ell^2_B\right)\left(eL/hc\right)\delta A_z$. Relating $\delta A_z$ to the electric field, we find
\begin{equation}
\Delta N_+-\Delta N_-=\frac{1}{2\pi^2}\frac{e^2}{\hbar^2 c}\int dt
\int d^3\br \;\bE\cdot \bB.
\end{equation}
This expression for the non-conservation of the total ``chiral'' charge is a direct consequence of the Adler-Bell-Jackiw anomaly.

\section{Many-body interactions}
\label{sec: many body}

In all condensed matter applications, the velocity of the massless Dirac particles, $v_F$, is much smaller than the speed of light in vacuum, $c$. This difference is important when many-body interactions are considered, and therefore, from now on, we shall intentionally distinguish between $v_F$ and $c$.

In a 2D semi-metal such as graphene, we can imagine integrating out all high-energy electronic modes {\it outside} of a finite energy interval about the Dirac point. The Fermi level is assumed to be close to the energy of the Dirac point. Since none of the gapless modes have been integrated out, there can be no non-analytic terms generated at long wavelengths, and in particular no screening of the $1/r$ electron-electron interaction whose 2D Fourier transform is, of course, non-analytic in momentum. Indeed, the long distance tail of the bare electron-electron interactions falls off as $e^2/\left(4\pi \eps_d \; r\right)$, where $\eps_d$ is the dielectric constant of the 3D medium in which the graphene sheet has been embedded. At long distances, $\eps_d$ is {\it independent} of the screening within the graphene sheet coming from the core carbon electrons. This can be shown by solving an elementary electrostatic problem of a point charge inserted in the middle of an infinite dielectric slab of finite thickness placed in a 3D medium with a dielectric constant $\eps_d$ \cite{Smythe1950}\cite{Emelyanenko2008}\cite{Wehling2011}. At distances much greater than the thickness of the slab, the Coulomb field within the slab is entirely determined by $\eps_d$.
A finite on-site Hubbard-like interaction is usually taken to model the very short distance repulsion.

What then are the consequences of such electron-electron interactions if the Dirac point coincides with the Fermi level?
The importance of each of the terms can be determined by dimensional analysis: in 2D, the Dirac field scales as an inverse length and therefore the short distance (contact) coupling $g$, multiplying four Dirac fields, has dimensions of length. In any perturbative series expansion, each power of $g$ must be accompanied by a power of an inverse length to maintain the correct dimensions of a physical quantity that is being computed. Since it is critical, the only lengthscales in the problem are associated with finite temperature, i.e. the thermal length $\hbar v_F/k_BT$, or the wavelength (frequency) of the external perturbation. As such length scales become very long, each term in the perturbative series in $g$ becomes small and we expect the series to converge. In the parlance of critical phenomena, the short range interaction is perturbatively irrelevant at the non-interacting (Gaussian) fixed point (see e.g. Ref.\cite{Herbut2006}). Therefore, while there can be finite modifications of the Fermi velocity or of the overlap of the true (dressed) quasiparticle with the free electron wave function, the asymptotic infrared properties of the model must be identical to the non-interacting Dirac problem\cite{GiulianiMastropietro2010,GiulianiMastropietro2009}.

Using a similar analysis for the $1/r$ tail of the non-retarded Coulomb interaction, one finds that $e^2/\left(\eps_d \hbar v_F\right)$ is dimensionless. Despite the superficial similarity with the 3+1D QED fine structure constant $e^2/\hbar c$, the physics here is different. First of all, the charge, being a coefficient of a non-analytic term in the Hamiltonian, does not renormalize when high energy modes are progressively integrated out\cite{Fisher1990}\cite{Herbut2001}. Any renormalization group flow of the dimensionless coupling $e^2/\left(\eps_d \hbar v_F\right)$ must therefore originate in the flow of $v_F$, which is no longer fixed by the Lorentz invariance because such symmetry is violated by the instantaneous Coulomb interaction. Detailed perturbative calculations reveal\cite{Gonzalez1994} that $v_F$ grows to infinity logarithmically at long distances thereby shrinking $e^2/\left(\eps_d \hbar v_F\right)$. Physically, however, $v_F$ cannot exceed the speed of light $c$. Instead, once the retarded form of the electron-electron interaction is properly included via an exchange of a (3D) photon, the flow of $v_F$ saturates at $c$. The resulting theory is quite fascinating, in that the 2D massless Dirac Fermions and the 3D photons propagate with the speed of light and, unlike in 3+1D QED, the coupling $e^2/\hbar c$ remains finite in the infra-red\cite{Gonzalez1994}. Unfortunately, since the flow of $v_F$ is only logarithmic, and since initially there is a large disparity in the values of $v_F$ and $c$, such a fixed point is practically unobservable. Instead, in practice, the physics is at best given by the crossover regime in which $v_F$ increases, but never to values comparable to $c$.

 The $1/r$ Coulomb interaction induced enhancement of the Fermi velocity is expected to lead to a suppression of the low temperature specific heat {\it below} its non-interacting value \cite{VafekPRL2007}, as well as other thermodynamic quantities \cite{SheehySchmalianPRL2007}. Interestingly, the suppression of the single particle density of states does not lead to a suppression of the ac conductivity; in the non-interacting limit it takes a (frequency independent) value $\sigma_0=Ne^2/16\hbar$ where $N$ is the number of the 2-component ``flavors''.
Again, the reason is the enhancement of the velocity: loosely speaking, while there are fewer excitations at low energy, those that are left have a higher velocity and therefore carry a larger electrical current.
The expression \cite{HerbutJuricicVafekPRL2008} for the low frequency ac conductivity has the form $\sigma(\omega)=\sigma_0\left(1+Ce^2/(\hbar v_F+  \frac{e^2}{4}\log \frac{v_F\Lambda}{\omega})\right)$, where $\Lambda$ is a large momentum cutoff. In the limit $\omega\rightarrow 0$, the correction to the non-interacting value is seen to vanish\cite{SheehySchmalianPRL2007}\cite{HerbutJuricicVafekPRL2008}\cite{MishchenkoEPL2008}. The value of the (positive) constant $C$ in this expression has been a subject of debate as it seems to depend on the details of the UV regularization procedure\cite{HerbutJuricicVafekPRL2008}\cite{MishchenkoEPL2008}\cite{SheehySchmalianPRB2009}
\cite{JuricicVafekHerbutPRB2010}\cite{SodemannPRB2012}\cite{KotovPRB2008}. Recently, the calculation of $C$ within a honeycomb tight-binding model \cite{Rosenstein2013}, which provides a physical regularization of the short distance physics, found $C=11/6-\pi/2\approx 0.26$; this value was also obtained within a continuum Dirac formulation using dimensional regularization \cite{JuricicVafekHerbutPRB2010} by working in $2-\eps$ space dimensions, and eventually setting $\eps=0$.

Increasing the strength of the electron-electron interactions, while holding the kinetic energy fixed, is expected to cause a quantum phase transition into an insulating state with a spontaneously generated mass for the Dirac Fermions \cite{Khveshchenko2001}\cite{Leal2004}. Since, as we just argued, weak interactions are irrelevant at long distances, such transition must happen at strong coupling, making it hard to control within a purely Fermionic theory. The full phase diagram also depends on the details of the interaction and is difficult to determine reliably using analytical methods. However, if one {\it assumes} that there is a direct continuous quantum phase transition between the semi-metallic phase at weak coupling and a known broken-symmetry strong coupling phase, say an anti-ferromagnetic insulator, then the critical theory can be argued to take the form of massless Dirac Fermions Yukawa-like coupled to the self-interacting order parameter bosonic field \cite{HerbutJuricicVafek2009}. The advantage of this formulation is that the upper critical (spatial) dimension is 3, and therefore such theory can be studied in $3-\epsilon$ space dimensions within a controlled $\epsilon$-expansion, eventually extrapolating to 2 space dimensions by setting $\epsilon=1$. The transition thus found is indeed continuous and governed by a fixed point at finite Yukawa and quartic bosonic couplings. To leading order in $\epsilon$, the critical exponents have been determined\cite{HerbutJuricicVafek2009}; for the semi-metal to the antiferromagnetic insulator quantum phase transition, the correlation length exponent $\nu=0.882$ and the bosonic anomalous dimension $\eta_b=0.8$. Since the dynamical critical exponent has been found to be $z=1$, these values imply that the order parameter vanishes at the transition as $|u-u_c|^{\beta}$ with the exponent $\beta=0.794$; here $u_c$ is a critical interaction. The $1/r$ Coulomb interaction has been found to be irrelevant at this fixed point.

Given that at half-filling the theory does not suffer from the Fermion sign problem, a very promising theoretical approach in this regard is numerical. The Hubbard model on the honeycomb lattice, with the nearest neighbor hopping energy $t$ and the repulsive on-site interaction $U$, has been studied using quantum Monte Carlo methods \cite{SorellaTosatti1992}\cite{Paiva2005}\cite{Meng2010}\cite{Sorella2012}. Recent simulations on cluster sizes of up to 2592 sites show strong indications of a direct continuous phase transition at $U/t\approx 3.869\pm 0.013$ between the (Dirac) semi-metal and the anti-ferromagnetic insulator\cite{Sorella2012}, disfavouring earlier claims\cite{Meng2010} on the existence of a spin liquid phase for intermediate values of couplings $3.4\lesssim U/t\lesssim 4.3$ using smaller cluster sizes of up to 648 sites. The critical exponent $\beta= 0.8\pm0.04$ extracted in Ref.\cite{Sorella2012} is in excellent agreement with the value obtained using the analytic Yukawa-like theory\cite{HerbutJuricicVafek2009}. In subsequent numerical simulations, the anti-ferromagnetic order parameter has been pinned by introducing a local symmetry breaking field\cite{AssaadHerbut2013}. The resulting induced local order parameter far from the pinning center was then `measured'. This procedure resulted in an improved resolution, confirming a continuous quantum phase transition between the semi-metallic and the insulating anti-ferromagnetic states. The single particle gap was found to track the staggered magnetization, while the critical exponents obtained from finite size scaling agree with those obtained to leading order in $\eps$-expansion \cite{HerbutJuricicVafek2009}.

The $1/r$ Coulomb interaction can also be simulated efficiently without the Fermion sign problem using a hybrid Monte Carlo algorithm \cite{DrutLahtePRL2009} using either staggered Fermions \cite{DrutLahtePRL2009}\cite{DrutLahtePRB2009a}\cite{DrutLahtePRB2009b} or, preferentially, directly on a honeycomb tight-binding lattice\cite{Brower2011a}\cite{Brower2012a}\cite{Buividovich2012a}\cite{Buividovich2012b}\cite{Ulybyshev2013}. The critical strength of the interaction necessary to achieve a quantum phase transition into an insulating state seems to depend on the details of the short distance part of the repulsion. Moreover, the system sizes studied numerically \cite{Ulybyshev2013} may be too small to explore the unscreened long distance tail of the $1/r$ interactions and to therefore unambiguously establish theoretically whether suspended monolayer graphene should be insulating.
It is worth pointing out here that experiments on the suspended high purity monolayer graphene samples show no sign of spontaneous symmetry breaking and would thus place it on the semi-metallic side.

\section{Applications to various physical systems}

\subsection{Graphene}
It is interesting to consider the massless Dirac Fermions in graphene\cite{GeimMacDonald2007} within the perspective outlined above.
Pure symmetry arguments are a powerful tool in this regard; our goal is to carry out such arguments in full detail in this section
in order to illustrate their utility.
Assuming a perfectly flat, $sp^2$ hybridized carbon sheet, the relevant atomic orbitals forming both the conduction and the valence bands are the carbon $2p_z$ orbitals\cite{GeimMacDonald2007}\cite{CastroNetoRMP2009}. A good variational ansatz for $u_{1\bk}(\br)$ would be $\sum_{\bR}e^{i\bk\cdot\bR}\phi_{p_z}(\br-\bR-\frac{1}{2}\boldsymbol\delta)$, where $\phi_{p_z}(\br)$ is a L\"{o}wdin orbital\footnote{The L\"{o}wdin orbitals, as used by Slater and Koster\cite{SlaterKoster1954}, are linear combinations of the atomic orbitals that are orthogonal to each other on different sites.} with the same symmetry as the atomic $p_z$ orbital\cite{SlaterKoster1954}. The exact form of the L\"{o}wdin orbital is unimportant for us now, its symmetry is what matters. In an idealized situation, without externally imposed strains or any other lattice distortions, the set of vectors $\bR$ could be chosen to span the triangular sublattice of the graphene honeycomb lattice: $m\bR_1+n\bR_2$ with $\bR_1=\sqrt{3}\hat{x}$, $\bR_2=\frac{1}{2}\bR_1+\frac{3}{2}a\hat{y}$, and $m,n$ are integers. The basis vector $\boldsymbol\delta=\frac{\sqrt{3}}{2}a\hat{x}+\frac{1}{2}a\hat{y}$.
Note that this Bloch state is manifestly periodic in $\bk$. Similarly, we can choose $v_{1\bk}(\br)$ as $\sum_{\bR}e^{i\bk\cdot\bR}\phi_{p_z}(\br-\bR+\frac{1}{2}\boldsymbol\delta)$. This physically motivated choice, along with $u_{2\bk}(\br)=v_{2\bk}(\br)=0$, defines our four basis states used to construct the Eq.(\ref{eq:Heff Kramer's}).

A flat graphene sheet is invariant under the mirror reflection about the plane of the lattice which further constrains $H(\bk)$. Such operation reverses the in-plane components of the electron spin --- an axial vector ---  and leaves the perpendicular component unchanged, thus acting on the spin state as a $\pi$-rotation about the axis perpendicular to the graphene sheet. Additionally, the $p_z$ orbitals are odd under the mirror reflection.
Therefore, the effective Hamiltonian in Eq.\ref{eq:Heff Kramer's} is constrained to satisfy $1\otimes\sigma_3\; H(\bk)\;1\otimes\sigma_3=H(\bk)$ for any in-plane $\bk$. This forces $g_4=g_5=0$ in the Eq.(\ref{eq:Heff Kramer's}). Because the remaining three $g_{j}$'s are in general non-zero, we see that with only two components of $\bk$ we cannot find simultaneous zeros of three independent functions. Therefore, in the absence of any other symmetry, we should expect level repulsion.

We can find the location of the Dirac points by taking into account additional symmetries.
The space inversion symmetry, say about the center of the honeycomb plaquette, requires $\tau_1\otimes1\; H(-\bk)\; \tau_1\otimes1=H(\bk)$.
This forces $g_{1}(\bk)$ and $g_{3}(\bk)$ to be odd under $\bk\rightarrow -\bk$ and $g_2(\bk)$ to be even.
If the lattice also has a threefold symmetry axis perpendicular to the sheet and passing through the plaquette center, then $g_2$ and $g_3$ must vanish at the two inequivalent points $\bk=\pm\bK=\pm \frac{4\pi}{3\sqrt{3}a}\hat{x}$, as well as, of course, all points equivalent to $\pm\bK$ by periodicity in the momentum space.
This follows from our formalism when we note that the effect of the $\frac{2\pi}{3}$ rotation, induced on our wavefunctions by the operator $e^{-i\frac{2\pi}{3\hbar}\hat{L}_z}e^{-i\frac{\pi}{3}\sigma_3}$, affects our four basis states as $e^{i\phi\tau_3\otimes\sigma_3}e^{-i\frac{\pi}{3}1\otimes\sigma_3}$, where $\phi=\bk'\cdot\bR_1$ and $\bk'$ is the result of rotating $\bk$ counter-clockwise by $120^\circ$.
Then, the identity $e^{i\phi\tau_3\otimes\sigma_3} H(\bk') e^{-i\phi\tau_3\otimes\sigma_3}=H(\bk)$ evaluated at $\bk=\pm\bK$ immediately leads to $g_2(\pm\bK)=g_3(\pm\bK)=0$.
Interestingly, $g_1$ is finite at $\pm \bK$ with vanishing derivatives, although if we also assumed spin $SU(2)$ symmetry, which allows us to flip the spins using $\tau_1\otimes\sigma_1$, then $g_1$ would vanish as well. In such case, irrespective of the microscopic details of the full Hamiltonian, the two bands must touch at $\pm\bK$.

The Dirac particles of graphene therefore live at $\pm \bK$.
Strictly speaking, they are not quite massless because of non-zero spin-orbit coupling which makes $g_1(\bk)$ finite. Such a term has been introduced by Kane and Mele\cite{KaneMele2005}. However, this term is very small in planar graphene structures, because the carbon atom is light and because graphene has a reflection symmetry about the vertical plane passing through the nearest neighbor bond\cite{YaoPRB2007}\cite{Min2006}.
There is therefore only a negligibly small Dirac mass at $\bK$ of order $10^{-3}$meV.

Expanding  $H\left(\pm\bK+\delta\bk\right)$ to first order in $\delta\bk$ we find
\begin{eqnarray}\label{eq:Heff graphene}
H_{eff}=\pm m_{QSH} \tau_3\otimes1\pm \hbar v_F\delta k_{\parallel}\tau_1\otimes1+ \hbar v_F\delta k_{\perp}\tau_2\otimes\sigma_3,
\end{eqnarray}
where the 3-fold rotational symmetry guarantees that the $\delta k_{\parallel}$ and $\delta k_{\perp}$ are two mutually orthogonal projections of $\delta \bk$.
In the coordinate system we have adopted, the mirror reflection symmetry about the $x-z$ plane forces $\delta k_{\parallel}=\frac{1}{2}\delta k_x+\frac{\sqrt{3}}{2}\delta k_y$ and $\delta k_{\perp}=-\frac{\sqrt{3}}{2}\delta k_x+\frac{1}{2}\delta k_y$.
At energy scales much smaller that $m_{QSH}$, this Hamiltonian describes the quantum spin Hall state: a gapped phase with counter-propagating edge states\cite{KaneMele2005}.
Due to the smallness of $m_{QSH}$ in graphene, for all practical purposes we can set it to zero.
The particle hole asymmetry, which arises from the $\delta\bk^2$ dependence of $g_1$, is also small in that it guarantees that the Fermi level can in principle be tuned to the Dirac point without the appearance of additional Fermi surfaces. The value for the Fermi velocity, $v_F\approx 10^6m/s$, can be obtained from approximate first principle calculations or from experiments.

\subsubsection{Coupling to external fields}
Perhaps the greatest utility of the Dirac-like equation (\ref{eq:Heff graphene}) is its ability to capture {\it both} the kinematics of the low energy excitations {\it and} their dynamics when subjected to external, or internal, fields. The former are of course the experimental tool of choice in studying the system.

In our theoretical description, we are tempted to minimally couple the external vector potential $\bA(\br)$, associated with the perpendicular magnetic field $\bB(\br)=\nabla\times\bA(\br)$, and scalar potential associated with either an applied electric field or to the field induced by impurities. While some care must be applied since we are working with a Bloch basis whose periodic part changes with $\bk$, to the order in $\delta\bk$ that the Eq.\ref{eq:Heff graphene} has been written, we are actually allowed to perform such minimal substitution\cite{LuttingerKohn1955}\cite{DiVincenzoMele1984}. Therefore, as long as the fields are sufficiently weakly varying in space, or for the uniform magnetic field as long as the magnetic length $\sqrt{\hbar c/eB}$ is much longer than the lattice spacing, we have
\begin{eqnarray}\label{eq:Heff graphene ext pots}
H_{eff}=
\pm v_F\left(p_\parallel-\frac{e}{c}A_\parallel(\br)\right)\tau_1\otimes1+ v_F\left(p_\perp-\frac{e}{c}A_\perp(\br)\right)\tau_2\otimes\sigma_3+U(\br)1_4+H_Z.
\end{eqnarray}
where the Zeeman term is $H_Z=\frac{1}{2}g\mu_B \left(B_x\tau_1\otimes\sigma_1 + B_y\tau_1\otimes\sigma_2 +B_z 1\otimes\sigma_3\right)$.
The above Hamiltonian governs the behavior of graphene in an external magnetic field. The resulting Landau level structure has been directly observed in scanning tunneling spectroscopy\cite{Miller2009}\cite{LiPRL2009}\cite{Luican2011}. Its utility in understanding the experiments on graphene hetero-junctions has been reviewed in Ref.\cite{YoungKim2011}.
The Schwinger mechanism, discussed in Section\ \ref{sec:external perturbations}, has been experimentally tested in Ref.\cite{Vandecasteele}.
$H_{eff}$ can also accommodate a time dependence of external potentials, important for interpreting the optical\cite{NairScience2008} or infra-red spectroscopy measurements of graphene\cite{LiNatPhys2008}.
The enhancement of the Fermi velocity, which, as discussed in Section \ref{sec: many body}, is a signature of electron-electron interactions, have been reported in Ref.\cite{Elias2011}, with no signs of gap opening at the Dirac point.
The effects of strain, as an effective potential in $H_{eff}$, are discussed in Refs.\cite{Guinea2010}\cite{Levy2010}\cite{Vozmediano2010}.
By and large, realistic impurity potentials in graphene cannot be treated in linear response theory\cite{DiVincenzoMele1984}\cite{YangWangScience2013}; the review of transport effects can be found in Ref.\cite{DasSarmaRMP2011}.

\subsection{Surface states of a 3D topological insulator}
An example of a 3D topological insulator\cite{FuKaneMele2007}\cite{FuKanePRB2007}\cite{MooreBalentsPRB2007}\cite{RoyPRB2009} is $Bi_2Se_3$\cite{HZhangNatPhys2009}\cite{CXLiuPRB2010}\cite{HasanMoore2011}. Its excitation spectrum is gapped in the 3D bulk, but its 2D surfaces accommodate gapless excitations which carry electrical charge, conduct electricity, and the dispersion of the surface excitations obeys massless Dirac equation. Unfortunately, presently the actual material suffers from imperfections causing finite bulk conductivity, a complication which we will largely overlook in this review.

The electronic configuration of $Bi$ is $6s^2 6p^3$ and of $Se$ is $4s^2 4p^4$. Since the $p$-shells of $Se$ lie $\sim2.5eV$ below $Bi$\cite{HarrisonBook1989}, a naive valence count would suggest that the two $Bi$ atoms donate six of their valence $p$-electrons to fill the $p$-shell of $Se$. We would therefore incorrectly conclude that the system is a simple, or trivial, insulator with a fully filled $Se$-like $p$-band and empty $Bi$-like conduction band, perhaps with an appreciable band gap. Interestingly, the strong spin-orbit coupling causes a ``band inversion''\cite{HZhangNatPhys2009}\cite{CXLiuPRB2010} near the $\Gamma$-point (the origin of the Brillouin zone), where the $Bi$-like states lie below the $Se$-like states. Because the rhombohedral crystal structure of $Bi_2Se_3$ has a center of inversion, the exact Bloch eigenstates must be either even or odd under space inversion at the crystal momenta which map onto themselves under time reversal, modulo a reciprocal lattice vector, i.e., $\bk=-\bk+\bG$. Clearly, $\Gamma$ is such a point. As shown by Fu and Kane \cite{FuKanePRB2007}, a sufficient condition for a band insulator with a center of inversion to be a 3D topological insulator is if such band inversion happens at an odd number of time reversal invariant points. More precisely, the system is a 3D topological insulator if the product of the parity eigenvalues of the occupied bands at the time reversal invariant $\bk$-points is odd, with the understanding that we count the parity eigenvalue of only one of the members of the Kramers pair.
This is indeed what happens within a more realistic band structure calculation\cite{HZhangNatPhys2009} \cite{CXLiuPRB2010} of $Bi_2Se_3$. At the $\Gamma$ point --- but not at the other time reversal invariant $\bk$-points --- the parity even combination of the $p_z$-like $Bi$ states are spin-orbit coupled to the more energetic $p_x\pm ip_y$-like $Bi$ states, and get pushed below the parity odd combination of the $Se$ $p_z$-like and $p_x\pm ip_y$-like states.

The Eq.\ref{eq:Heff Kramer's} must describe the dispersion near the $\Gamma$ point inside the bulk of the 3D system. This can be seen explicitly if we choose $u_{1\bk}(\br)$ to be predominantly made of the parity even combination of $Bi$ $p_z$-like orbitals and $-u_{2\bk}(\br)$ of the $Bi$ $p_x+ip_y$-like orbitals; i.e., the states which are mixed due to the spin-orbit interaction. Similarly, for the proximate band, we should have $v_{1\bk}(\br)$ made predominantly of the parity odd combination of the $Se$ $p_z$-like orbitals, and $-v_{2\bk}(\br)$ of $Se$ $p_x+ip_y$-like orbitals\cite{CXLiuPRB2010}.
Then, up to the quadratic order in deviation from the $\Gamma$ point, $g_1(\bk)=M_0+M_1k^2_z+M_2\left(k^2_{x}+k^2_{y}\right)$ with $M_0<0$ and $M_{1,2}>0$. No $\bk$-odd terms are allowed here because the states are of definite parity. Note that because $M_0$ is negative, in the immediate vicinity of the $\Gamma$ point the $Bi$-like states lie below the $Se$-like states. At higher $\bk$, we revert to the expected band ordering.
For the other terms in the Eq.\ref{eq:Heff Kramer's}, $g_2(\bk)=0$ to linear order in $\bk$, due to additional 3-fold rotational symmetry; it is non-zero when we include terms up to order $k^3$, since the $\bk$-cubic invariant exists. The remaining terms must be $\bk$ odd, because they couple opposite parity states: to linear order then, $g_3(\bk)=B_0k_z$, $g_4(\bk)=-A_0k_x$, and $g_5(\bk)=-A_0k_y$, where $A_0\gtrsim B_0>0$. The particle-hole symmetry breaking term $f(\bk)$ is also finite, but since its presence leads to qualitatively same conclusions, it will be ignored\cite{CXLiuPRB2010}.

Since $g_1(\bk)$ is finite at $\Gamma$, which in this approximation is the only place where $g_3$, $g_4$, and $g_5$ vanish, the spectrum in the bulk is of course gapped. However, the surface is gapless. To see this explicitly\cite{HZhangNatPhys2009,CXLiuPRB2010}, consider a semi-infinite interface in the $x-y$ plane, set $k_x=k_y=0$, and construct evanescent zero energy states along the $z$-direction. There are always two such normalizable states, which can be used as a basis for the low energy subspace. The effective surface Hamiltonian for small $k_x$ and $k_y$ can be obtained by sandwiching the bulk Hamiltonian between these two states. For macroscopically thick material, we can ignore the exponentially small overlap between the surface states, and we find
$H_{surf}=\pm A_0\left(k_x\sigma_y-k_y\sigma_x\right)$,
where the top sign is for the top surface, $z=L$, and the bottom sign for the bottom surface $z=-L$.
A similar procedure along the right, $y=L$, and left, $y=-L$, surfaces leads to $H_{surf}=\pm \left(B_0k_z\sigma_x+A_0k_x\sigma_z\right)$; the effective Hamiltonians are simply related to each other by space inversion.
In general,
\begin{eqnarray}
H_{surf}=\hat{\bn}'\cdot\left(\vec\sigma\times \bk'\right)
\end{eqnarray}
where $\hat{\bn}'$ is obtained by rotating the normal to the surface, $\hat{\bn}$, by $180^\circ$ about the z-axis, and $\bk'=(-A_0k_x,-A_0k_y,B_0k_z)$.
We thus arrive at an equation for massless, anisotropic, Dirac particles. However, unlike in graphene which has four ``flavors'', the surface of the 3D topological insulator can support a single flavor.

\subsubsection{Coupling to external fields, interaction and disorder effects}
The existence of a single Dirac flavor on the surface of the 3D topological insulator has important consequences for robustness of the surface states towards impurity disorder. The states at $\bk$ and at $-\bk$ have opposite spin, leading to the suppression of back scattering\cite{TAndo1998a}\cite{TAndo1998b} and absence of localization for weak (scalar potential) disorder\cite{BardarsonPRL2007}\cite{NomuraPRL2007}\cite{LewenkopfPRB2008}. Theoretically, such a (non-interacting) system is always expected to display electrical conductivity which increases towards infinity as a logarithm of the system size.
Recall that in graphene with a pair of Dirac cones at $\bK$ and $-\bK$, such back scattering is always present and therefore weak localization is expected to eventually set in\cite{McCannPRL2006}\cite{AleinerEfetovPRL2006}, although for smooth impurity potentials, it may be very small\cite{ShonAndo1998}\cite{MuccioloReview2010}.

Recent numerical study\cite{SchubertPRBr2012} of a topologically non-trivial 3D lattice model --- with random on-site energy intentionally placed only on the surface of the 3D system --- indicates, that the effective continuum description with Dirac particles scattered by a scalar potential holds if the disorder strength is much weaker than the bulk gap ($\sim0.3eV$ in $Bi_2Se_3$). The assertion is based on identification of Dirac-like features in a momentum resolved spectral function, even when the translational symmetry of the lattice is broken by disorder. As the typical disorder strength increases beyond the 3D bulk gap value, the surface states appear diffusive. For even larger disorder strength, the outermost surface states are localized, but {\it weakly} disordered Dirac-like states reappear directly beneath it. Apparently, for large surface disorder, an interface between a strongly localized Anderson insulator and a topological insulator is formed\cite{SchubertPRBr2012}.
As such calculations were performed on finite size systems, which are too small to detect an Anderson localization transition, it is presently impossible to conclude whether there is a true phase transition at zero temperature separating the weak, the moderate, and the strong disorder regimes.
The combined effects of scalar disorder and electron-electron (Coulomb) repulsion have been studied in Ref.\cite{OstrovskyPRL2010} using the continuum Dirac approximation. The authors argue that 3D topological insulators are different from graphene, and that the single Dirac flavor makes the system metallic with finite conductivity at zero temperature.
Transport properties of topological insulators have been reviewed in Ref.\cite{BardarsonMoore2012}.

\begin{figure}[t]
\centering
\begin{tabular}{cc}
\includegraphics[height=8cm]{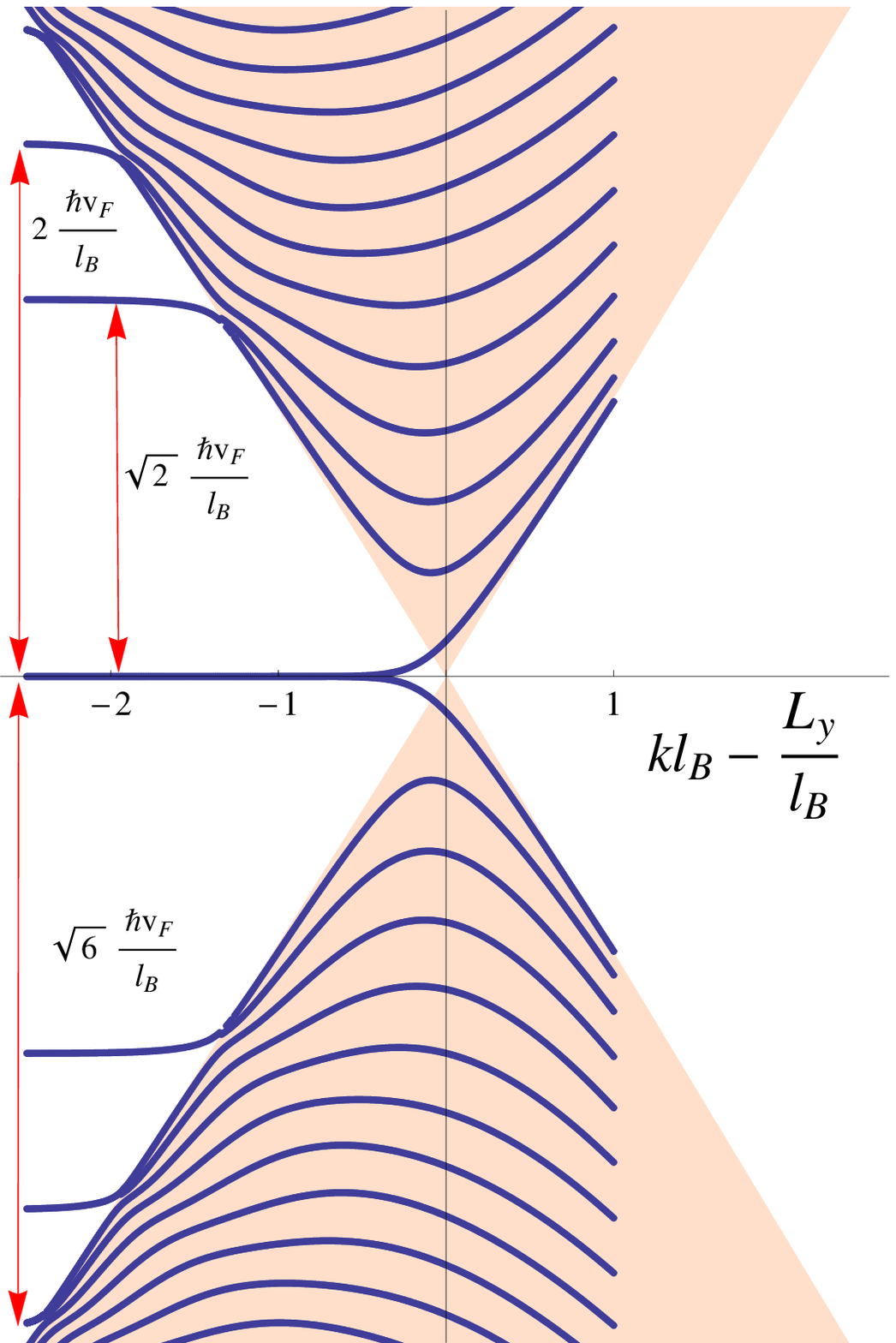} &
\begin{tabular}{c}
\includegraphics[height=4.5cm]{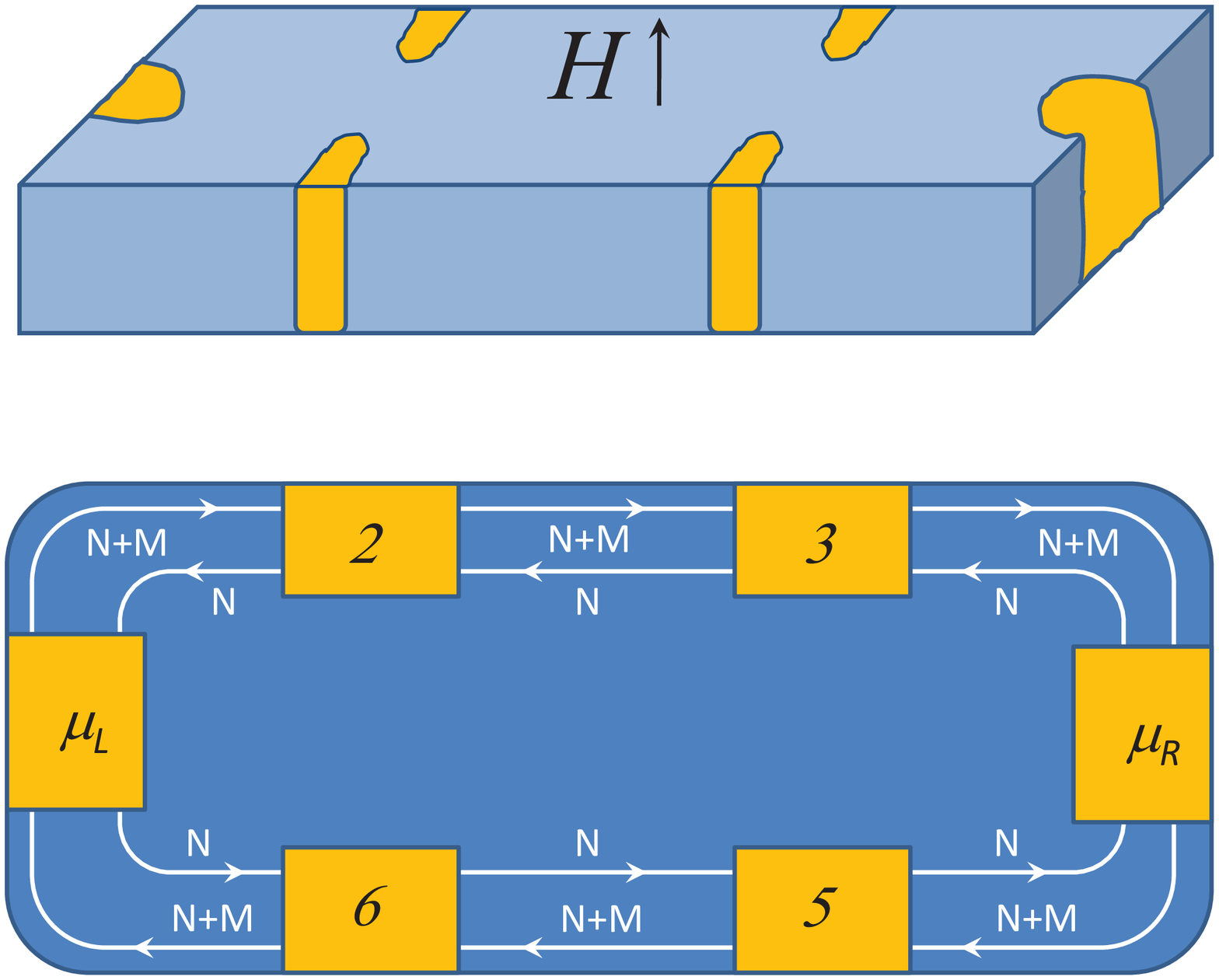} \\
(b)\\
\includegraphics[height=4.0cm]{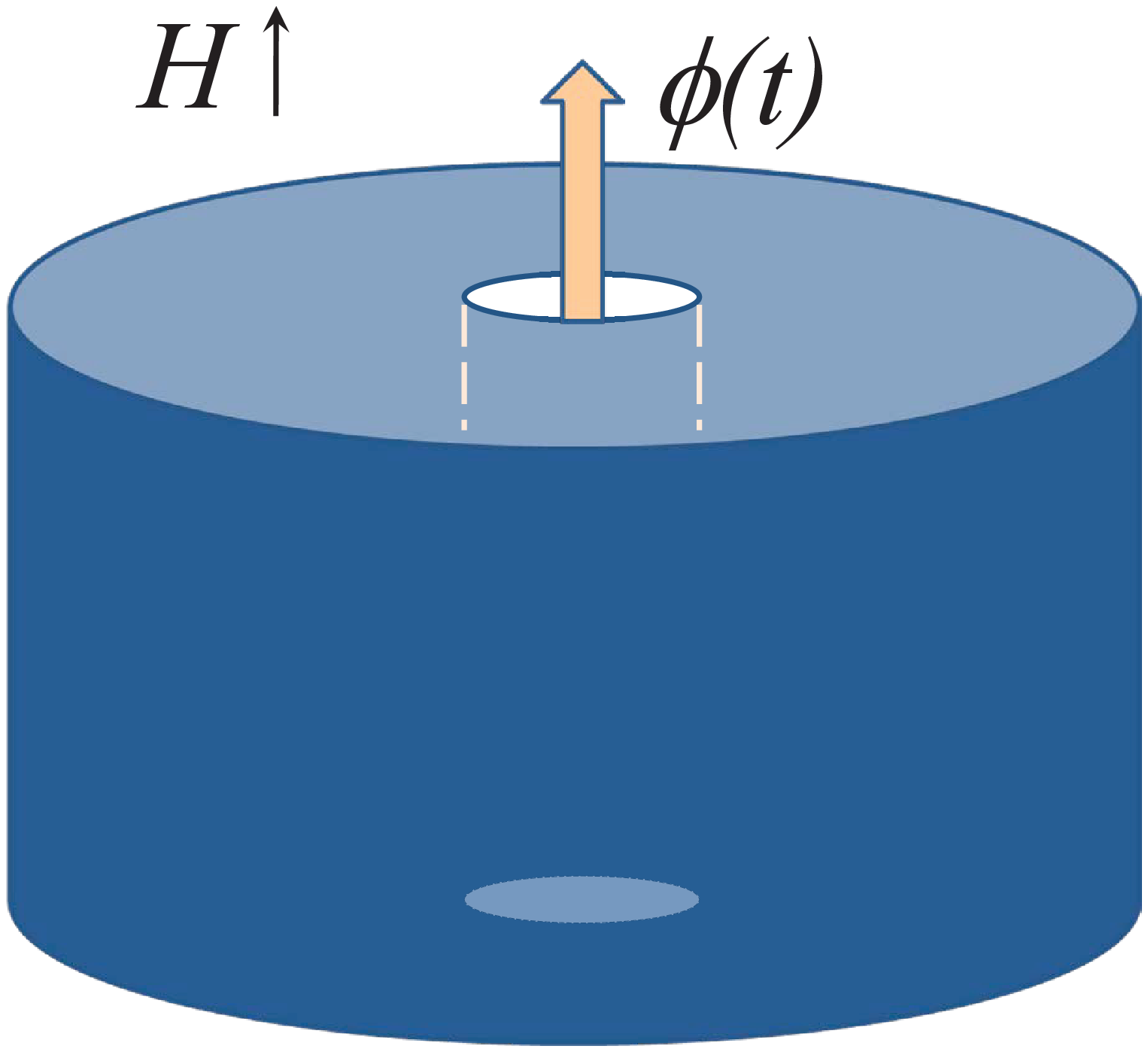} \\
(c)
\end{tabular} \\
(a)
\end{tabular}
\caption{(a) Electronic spectrum of a 3D topological insulator semi-infinite slab of finite thickness vs the "guiding center" coordinate. Far away from any edges, the spectrum exhibits the usual Dirac Landau level quantization, $E=\sqrt{n}\sqrt{2}\hbar v_F/\ell_B$, where the magnetic length is $\ell_B=\sqrt{\hbar c/eB}$, and $\hbar v_F=A_0\approx 3.3eV\AA$ for $Bi_2Se_3$. Every such Landau level is doubly degenerate.
If the Fermi level lies between the two Dirac Landau levels, the edge spectrum contains $M=2n+1$ chiral modes in addition to $2N$ non-chiral ones.
(b) Schematic of a Hall bar geometry in a 3D topological insulator. (c) Corbino geometry setup for measurements of quantum Hall conductivity.}
\label{fig:TIqhe}
\end{figure}

Because the electron spin is strongly coupled to its momentum, unlike in graphene, the Zeeman coupling to the external magnetic field does not lead to simple spin splitting. Rather, it opens up a gap, turning massless Dirac particles massive. To further illustrate the difference between the Dirac particles in a 3D topological insulator and graphene, consider now the situation in which the external uniform magnetic field is applied along the $z$-axis, and the field is sufficiently strong to quantize the orbital motion of the surface electrons. The equation describing the states on the top and the bottom surfaces is then
\begin{eqnarray}
\left[\pm v_F\left(\left(p_x+\frac{e}{c}By\right)\sigma_y-p_y\sigma_x\right)+g_z\mu_B B\sigma_z\right]\psi(x,y,\pm L)=E\psi(x,y,\pm L),
\end{eqnarray}
where $\hbar v_F=A_0$ and $g_z$ is the effective Lande g-factor. Indeed, the Zeeman coupling acts as a Dirac mass and does not lead to the usual splitting of the spin degenerate energy levels.
It is straightforward to find the eigenvalues of this operator provided we are sufficiently far from any edge. The resulting Landau level spectrum is
\begin{eqnarray}
E_n&=&\pm \sqrt{2A^2_0\left(\frac{eB}{\hbar c}\right)n+\left(g_z\mu_B B\right)^2},\;\;n=1,2,3,\ldots\;\\
E_0&=&g_z\mu_B B.
\end{eqnarray}
The physics in a quantizing magnetic field differs from graphene near the edge in another important way: the top and the bottom surfaces are coupled through the side surfaces. The applied magnetic field is parallel to the side surfaces and therefore there is no Landau quantization along this surface; even the Zeeman term does not open up a gap on the side surfaces, it merely shifts the momentum by a constant. Therefore, as the guiding center of the Landau levels approaches the edge, they start mixing into the continuum of the states in the side surfaces.
Fig.\ref{fig:TIqhe} shows the electronic spectrum of a 3D topological insulator semi-infinite slab of finite thickness vs. the ``guiding center'' coordinate.
Far away from any edges, the spectrum exhibits the usual Dirac Landau level quantization, $E=\sqrt{n}\sqrt{2}\hbar v_F/\ell_B$, where $\ell_B=\sqrt{\hbar c/eB}$ and for $Bi_2Se_3$, $v_F=A_0$.
Every such Landau level is doubly degenerate because the top and the bottom surfaces are assumed to be identical. Such degeneracy would be lifted if the inversion symmetry is broken by, say, a constant chemical potential difference between the top and the bottom surfaces.
As the guiding center coordinate approaches the right edge --- or the outer edge for the ``Corbino'' geometry --- the Landau level states merge with the plane-wave states from the vertical side surface. In the limit of very large thickness such plane-wave states form a Dirac continuum.

This poses interesting questions: how robust is the quantum Hall effect and how to measure it\cite{DHLeePRL2009}?
If the Fermi energy lies between the two Landau levels, the spectrum contains $M=2n+1$ chiral edge modes in addition to $2N$ non-chiral ones.
Clearly, in any Hall bar geometry the leads necessarily couple to the continuum of the states in the side surfaces, which present additional (unwanted) channels of conduction.
Assuming that the side modes equilibrate with each other and result in a finite conductivity, the chemical potential will drop smoothly between $\mu_R$ and $\mu_L$ along each edge, and no quantization of Hall conductance is expected\cite{DHLeePRL2009}\cite{VafekPRB2011}\cite{YYZhang2012}.
Interestingly, quantization of $\sigma_{xy}$ has been reported in a strained 70-nm-thick HgTe layer\cite{BrunePRL2011}, with a well developed plateau at $\nu=2$ and plateau-like features at $\nu=3$ and $4$. At the same time, the longitudinal resistance $R_{xx}$ measured at $50mK$ shows a suppression by few tens of percents, but it does not reach zero.
While this observation awaits a complete theoretical treatment, if the sample is thin then there are only a few non-chiral modes along the side surfaces which may get Anderson localized with sufficient side surface roughness, leaving only chiral modes at the edges.

On the other hand, measurement of $\sigma_{xy}$ in the Corbino geometry is expected to lead to quantization\cite{DHLeePRL2009}\cite{VafekPRB2011}. The idea\cite{VafekPRB2011} is to perform the analog of the Laughlin thought
experiment, experimentally realized in 2D electron gas heterostructures in Ref.\cite{Dolgopolov1992}. One measures
the amount of charge $\Delta Q$ transferred from the inner surface
to the outer surface in response to the induced EMF produced in the azimuthal direction by a slow change in the magnetic
flux $\Delta\varphi$ threading the sample. Then $\sigma_{xy}=-c\Delta Q/\Delta \varphi$. For $\sigma_{xy}=n\frac{e^2}{h}$, half of the charge travels through the top surface and the other half through the bottom surface.
An additional advantage of the Corbino setup is that any interaction-driven fractional
quantum Hall states formed by the surface electrons can in principle also be detected\cite{Dolgopolov1992}.

If the external electro-magnetic potentials are weak, the linear response theory is applicable. Naively, for a non-interacting system with a gap, we expect that at long wavelength and low frequency the response functions simply change, or renormalize, the dielectric constant and the magnetic permeability; after all, the system is a dielectric insulator. Interestingly, a 3D topological insulator gives rise to additional terms in the electro-magnetic response, some of which are analogous to axion electrodynamics\cite{Wilczek1987}\cite{XiHughesZhang2008}\cite{MFranz2008}\cite{EssinPRL2009}.

\subsection{$d_{x^2-y^2}$-wave superconductivity in copper oxides}

\begin{figure}[h!]
\centering
\begin{tabular}{c}
\includegraphics[width=0.7\textwidth]{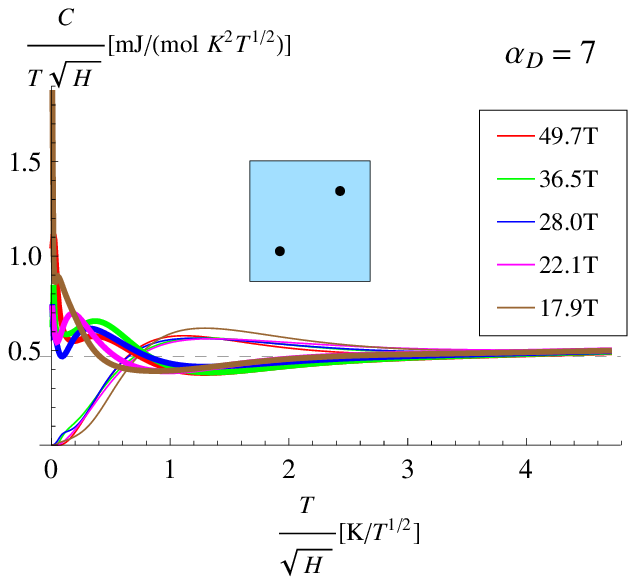} \\
(a)\\
\includegraphics[width=0.7\textwidth]{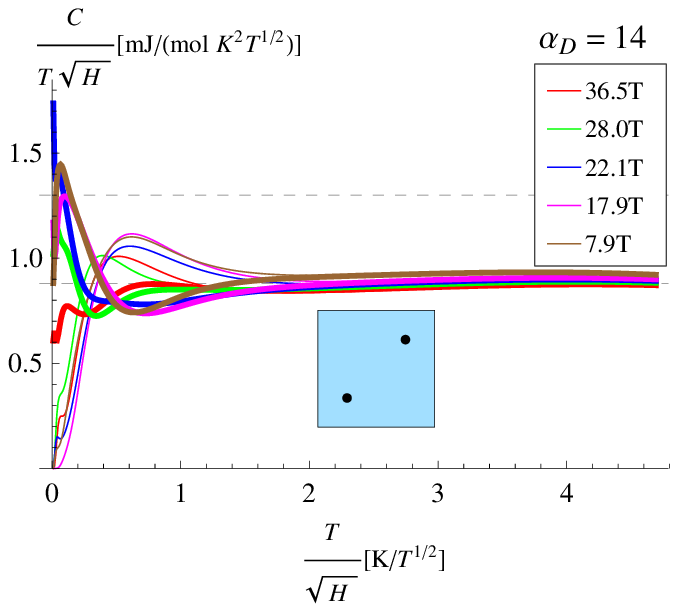}\\
(b)
\end{tabular}
\centering \caption{
Electronic contribution to the low temperature
specific heat of a $d_{x^2-y^2}$ superconductor in the vortex
state\cite{WangVafekunpublished}, scaled according to the Simon and
Lee scaling\cite{SimonLee1997}. The thick lines are with Zeeman term
included, the thin lines are without it. The electrons hop with the
nearest neighbor amplitude $t$ on a tight-binding lattice with a
lattice spacing $a=3.8\AA$. The chemical potential was set to
$\mu=0.297t$ corresponding to $15\%$ doping. The Fermi velocity
$v_F=2.15*10^5m/s$ was taken to agree with the photoemission
experiments on YBCO\cite{Damascelli2010} by setting $t=132$meV; the
Dirac cone anisotropy $\alpha_D=v_F/v_{\Delta}=7$ in panel (a) and
$\alpha_D=14$ in panel (b). Insets show the square vortex lattice
used. The dashed lines correspond to the values extracted
experimentally: (a) $\sim 0.47mJ/mol K^2 \sqrt{T}$ at $10\%$ doping
by Riggs {\it et.al.}\cite{Riggs2011}; (b) $\sim 0.87mJ/mol K^2
\sqrt{T}$ at $15\%$ doping by Moler {\it et.al.}\cite{Moler1994}
(lower dashed line) and $\sim 1.3mJ/mol K^2 \sqrt{T}$ at $15\%$
doping by Wang {\it et.al.}\cite{Wang2001} (higher dashed line); see
also \cite{Fisher2007}.} \label{fig:dwave specific heat}
\end{figure}
Low energy quasiparticles obeying the Dirac equation may also emerge as a consequence of a phase transition associated with the condensation of Cooper pairs.
The specific example which we consider here is the so called $d_{x^2-y^2}$ pairing which occurs in cuprate high temperature superconductors \cite{vanHarlingen1995}\cite{Kirtley2000}.
In these layered, quasi 2D, materials, one may focus on the electronic structure of a single CuO$_2$ layer. A simple effective Hamiltonian for this system is
\begin{eqnarray}H=\sum_{\bk,\sigma}\left(\eps_{\bk}-\mu\right)c^{\dagger}_{\sigma}(\bk)c_{\sigma}(\bk)+
\sum_{\bk}\left(\Delta_{\bk}c^{\dagger}_{\uparrow}(\bk)c^{\dagger}_{\downarrow}(-\bk)+h.c.\right),
\end{eqnarray}
where $\bk=(k_x,k_y)$. The normal state dispersion, given by $\eps_{\bk}$, describes a closed Fermi surface, centered around $(\pi,\pi)$, and equivalent points in momentum space. The anomalous self-energy, $\Delta_{\bk}$, must in principle be determined from a microscopic theory; since such theory is currently missing, one proceeds phenomenologically. Assuming time reversal symmetry, $\eps_{\bk}=\eps_{-\bk}$, and $\Delta_{\bk}$ can be chosen real. Since it transforms as $x^2-y^2$, it must change sign under a $90^\circ$ rotation and vanish along the Brillouin zone diagonals, where it intersects with the Fermi surface at four inequivalent points.
Weak orthorhombic distortions, such as in YBCO, move the points of intersection slightly away from the zone diagonals\cite{Kirtley2006}, but do not change the low energy physics in an important way.

The energy spectrum of the Fermionic quasiparticles can be obtained by solving the Heisenberg equation of motion for $c_{\uparrow}(\bk)$ and $c^{\dagger}_{\downarrow}(-\bk)$:
\begin{eqnarray}
i\hbar \frac{\partial}{\partial t}\left(\begin{array}{c} c_{\uparrow}(\bk)\\ c^{\dagger}_{\downarrow}(-\bk)\end{array}\right)=
\left(\begin{array}{cc} \eps_{\bk}-\mu & \Delta_{\bk} \\
\Delta_{\bk} & -\eps_{\bk}+\mu
\end{array}\right)
\left(\begin{array}{c} c_{\uparrow}(\bk)\\ c^{\dagger}_{\downarrow}(-\bk)\end{array}\right),\label{eq:dwave Heisenberg}
\end{eqnarray}
finding $E(\bk)=\sqrt{\left(\eps_{\bk}-\mu\right)^2+\Delta^2_{\bk}}$.
Near the points of intersection between the Fermi surface and the zeros of $\Delta_{\bk}$, we may expand $\eps_{\bk}-\mu\approx \hbar v_Fk_{\perp}$ and $\Delta_{\bk}\approx \hbar v_\Delta k_{\parallel}$, where $k_{\perp}$ and $k_{\parallel}$ are the deviation perpendicular and parallel to the Fermi surface respectively. In the vicinity of such points, the above has the form of an anisotropic massless Dirac equation.

Interestingly, the Dirac node remains at zero energy even as the chemical potential, $\mu$, is varied. This is unlike in the previous examples, which involved Dirac particles in semiconductors, where $\mu$ must be fine tuned to coincide with the Dirac node, otherwise we have Fermi circles with finite density of states at zero energy. Furthermore, given that the system is a superconductor, the long range Coulomb interaction is screened.
Since the discovery of cuprates being $d_{x^2-y^2}$ superconductors, there has been a tremendous effort in trying to understand the role of various perturbations.
Here we focus on the question `How does such a system behave in an external magnetic field?'\cite{Volovik1993}\cite{WangMacDonald1995}\cite{SimonLee1997}\cite{YasuiKitaPRL1999}\cite{FranzTesanovic2000}
The first step towards answering this question is to recognize that the upper and the lower components of the `spinor' in Eq.\ref{eq:dwave Heisenberg} acquire an opposite phase under a U(1) charge gauge transformation, and therefore, an external magnetic field cannot couple minimally\cite{FranzTesanovic2000}\cite{MarinelliHalperinSimon2000}\cite{VafekMelikyanFranzTesanovic2001}\cite{VishwanthPRL2001}\cite{VafekMelikyanTesanovic2002}\cite{VishwanthPRB2002}.
Moreover, the pair potential must also be modified. In a mean-field calculation, it is computed self-consistently, with the solution depending on the value of the external magnetic field\cite{WangMacDonald1995}\cite{YasuiKitaPRL1999}. But even in the absence of a microscopic theory --- which may justify a self-consistent mean-field calculation --- we can establish this fact by noting, that near the transition temperature, the existence of the Ginzburg-Landau functional follows quite generally from the order parameter having the charge $2e$ and the transition being continuous. Given that in cuprates the magnetic penetration depth is much longer than the coherence length, for most of the magnetic field range the field penetrates in the form flux tubes and the order parameter phase winds by $2\pi$ near the core of each vortex.
Therefore, in the presence of the external magnetic field, the equation which generalizes Eq.\ref{eq:dwave Heisenberg} is
\begin{eqnarray}
i\hbar \frac{\partial}{\partial t}\left(\begin{array}{c} c_{\br\uparrow}\\ c^{\dagger}_{\br\downarrow}\end{array}\right)=
\sum_{\br'}\left(\begin{array}{cc} t_{\br\br'}-\mu_{\uparrow}\delta_{\br\br'} & \Delta_{\br\br'} \\
\Delta^*_{\br\br'} & -t^*_{\br\br'}+\mu_{\downarrow}\delta_{\br\br'}
\end{array}\right)
\left(\begin{array}{c} c_{\br'\uparrow}\\ c^{\dagger}_{\br'\downarrow}\end{array}\right),\label{eq:dwave Heisenberg mixed state}
\end{eqnarray}
where we assumed that the electrons hop on a square lattice given by $\br$, with a complex amplitude $t_{\br\br'}$. The phase of the complex singlet pair potential $\Delta_{\br\br'}$ winds by $2\pi$ when its center of mass coordinate encircles a vortex sufficiently far from the vortex core; its dependence on the relative coordinate has $d_{x^2-y^2}$ symmetry.

When the typical separation between vortices, set by $\sqrt{hc/eB}$, is much smaller than the penetration depth, the magnetic field inside is almost uniform.
Clearly, in such a case, the plane waves with the wave-number $\bk$ are no longer eigenstates of the kinetic energy operator. One may attempt to proceed by working with Landau levels, which, in the continuum limit of the above lattice model, are eigenstates of the kinetic energy operator for a uniform magnetic field\cite{YasuiKitaPRL1999}. However, the number of the Landau levels below the Fermi energy, as determined from the quantum oscillations experiments on the overdoped side of the phase diagram\cite{Hussey2003}\cite{Vignolle2008}, is of order $10^4$ at magnetic fields of 1Tesla, this number decreasing with $1/B$. The energy scale associated with the pair potential is approximately given by $\left(v_{\Delta}/v_F\right)E_F$, decreasing the number of Landau levels mixed by $\Delta_{\br\br'}$ by only one order of magnitude. Moreover, the resulting Hamiltonian matrix is dense, prohibiting the use of efficient algorithms for determining the eigenvalues of sparse matrices.

In the relevant magnetic field range $H_{c1}\ll H\ll H_{c2}$ a different approach was proposed by Franz and Tesanovic\cite{FranzTesanovic2000}, circumventing the use of the Landau level basis.
The idea is to map the problem onto an equivalent one but at zero {\it average} magnetic field, in which case the plane wave basis may be used. This can be accomplished by performing a singular gauge transformation, familiar in the context of the fractional quantum Hall effect. They then argued that the relevant low energy excitations reside in the vicinity of the Dirac nodal points, and that, in the continuum limit, the vortices together with the magnetic field act as an effective potential scattering the Dirac particles. As the magnetic field decreases so does the strength of the effective potential, making a natural connection with the zero field problem.
For each of the four massless Dirac particles, which were assumed to be decoupled\cite{SimonLee1997}, the combination $\bv_F\cdot\left(\frac{\hbar}{2}\nabla\phi-\frac{e}{c}\bA\right)$ entered the Dirac equation as an effective electrical potential, $\Phi$ \cite{FranzTesanovic2000}. Here $\nabla\times\bA=\bB$ and $\nabla\times\nabla\phi=2\pi\hat{z}\sum_{j}\delta(\br-\bR_j)$. The additional minus signs acquired by the quasiparticles upon encircling an odd number of vortices was encoded using a statistical $U(1)$ field, minimally coupled to the Dirac particles\cite{FranzTesanovic2000}. Such an approach provided an explicit method to (numerically) compute the scaling functions, whose existence was proposed earlier by Simon and Lee\cite{SimonLee1997}, as well as to test the validity of the semiclassical approach advanced by Volovik\cite{Volovik1993}.

In the vicinity of each vortex, the effective potential $\frac{\hbar}{2}\nabla\phi-\frac{e}{c}\bA$ grows with the inverse of the distance to the vortex. Since the kinetic energy of a massless Dirac particle also scales with inverse length, the vortices constitute a singular potential. It is therefore not obvious that the long wavelength expansion, which led to the effective Dirac description in the first place, can be directly applied. Indeed, in the continuum limit, one must carefully specify the boundary conditions at the vortex core by requiring that the effective Hamiltonian is a self-adjoint operator\cite{MelikyanTesanovic2007}. A choice of such, so called, self-adjoint extensions
should be determined by matching to a well regularized lattice theory. Unfortunately, so far, it has not been possible to determine their form. Since the choice is not unique, and since different physically allowable choices appear to lead to a qualitative difference in the low energy spectra (e.g. gapped or gapless), one is led to work with the lattice theory\cite{VafekMelikyanFranzTesanovic2001}\cite{VafekMelikyanTesanovic2002}\cite{VafekMelikyanPRL2006}\cite{MelikyanVafek2008}. The usual choice is to set $t_{\br\br'}=-t e^{-iA_{\br\br'}}$ where the magnetic flux, $\varphi$, through an elementary plaquette enters the Peierls factor via $A_{\br\br+\hat{\bx}}=-\pi y e\varphi/hc$ and $A_{\br\br+\hat{\by}}=\pi x e\varphi/hc$. The ansatz for the pairing term is $\Delta_{\br\br+\boldsymbol\delta}=\Delta_0 \eta_{\boldsymbol\delta}e^{i\theta_{\br\br+\boldsymbol\delta}}$, where the $d_{x^2-y^2}$-wave symmetry is encoded by $\eta_{\boldsymbol\delta}=+(-)$ for $\boldsymbol\delta\parallel\hat{\bx}(\hat{\by})$, and the vortex phase factor $e^{i\theta_{\br\br'}}=\left(e^{i\phi_{\br}}+e^{i\phi_{\br'}}\right)/|e^{i\phi_{\br}}+e^{i\phi_{\br'}}|$. This choice is motivated by its behavior in the long distance limit\cite{VafekMelikyanFranzTesanovic2001}\cite{MelikyanTesanovic2006}.

For a periodic vortex arrangement, and after the appropriate lattice version of the singular gauge transformation, one can take advantage of the Bloch theorem. The quasiparticle spectrum is then a function of a ``vortex crystal'' momentum $\bq$. It can be shown\cite{VafekMelikyanTesanovic2002} that if the vortex lattice has a center of inversion and if the Zeeman term is ignored, then for each eigenstate with an eigenvalue $E$ at $\bq$, there is a corresponding eigenstate with an eigenvalue $-E$ at the same $\bq$. Therefore, any zero energy state at a fixed $\bq$ must be at least two-fold degenerate. However, because our problem breaks time reversal symmetry, such a degeneracy can only be achieved by fine tuning an additional parameter besides the two components of $\bq$. Therefore, the quasiparticle spectrum of an inversion symmetric vortex lattice is in general gapped. The Zeeman term corresponds to a simple overall shift of the quasiparticle energy and does not destroy the avoided crossing, it simply moves it to a non-zero energy. Some further non-perturbative aspects of this problem have been discussed in Ref.\cite{VafekMelikyanPRL2006}.

In Figure \ref{fig:dwave specific heat} we show the quasiparticle contribution to the specific heat obtained by the numerical diagonalization of the resulting (sparse) Hamiltonian matrix for different values of magnetic field. The result is re-scaled according to the Simon and Lee scaling\cite{SimonLee1997}\cite{MelikyanTesanovic2006}. We see that in the mixed state of a $d_{x^2-y^2}$ superconductor, for $v_F/v_{\Delta}=7$ and $14$, increasing the magnetic field indeed increases the specific heat in an intermediate temperature window, in accord with the semiclassical prediction by Volovik\cite{Volovik1993}. At the lowest temperatures, however, there is a crossover into the quantum regime where the interference effects set in and the finite spectral gap rapidly decreases the specific heat. Note that the entropy at low $T$, i.e. $\int_0^T C(T')/T'dT'$, increases with an increasing magnetic field. Entropy must of course be conserved and independent of the magnetic field when $T\rightarrow\infty$; the effect comes from the transfer of the spectral weight from energies above $\sim\Delta_0$. It is similar to the effect discussed in the context of the Dirac particle in a periodic electrical potential whose average vanished, see Fig.\ref{fig:Diracelectricpotential}.

We see then, that despite being described by similar kinematics, there is a very important difference in the way the $d_{x^2-y^2}$-wave Dirac particles couple to the physical external magnetic field from the way the graphene or the 3D topological quasiparticles couple. In the latter case, the specific heat may oscillate with the field, but when averaged over few oscillations, its value is field independent. In the former case, it is the {\it average} value that increases with the external field.

\section{Weyl Semimetals}
\begin{quotation}ÒMy work always tried to unite the truth with the beautiful, but when I had to choose one or the other, I usually chose the beautiful.Ó  - Hermann Weyl (1885-1955)\end{quotation}
It has long been known that band touchings in three dimensions are very stable\cite{vonNeumann,Herring}, as described in Section 2. When the chemical potential lines up with the band touching points, and no other Fermi surfaces intersect it, a semimetal results.  The low energy dispersion of electrons then closely resembles the Weyl equation of particle physics, hence these semimetals have been termed Weyl semimetals\cite{Wan}. The generic form is shown in Equation 4. Initially, the Weyl equation was believed to describe neutrinos, which however had to be given up with with the discovery of neutrino mass. Thus, an experimental realization of a Weyl semimetal would be the first physical realization of this fundamental equation. Here we will briefly review  topological aspects of Weyl semimetals  and their possible realizations in solids. For simplicity, consider the following simplified form of Equation 4:
\begin{equation}
H_\pm = \pm v_F \left ( p_x\sigma_1 + p_y\sigma_2 +p_z\sigma_3 \right )
\label{eq:isotropicWeyl}
\end{equation}
where we have expanded about a pair of band touchings located at $k_\pm$ and have denoted $p=\hbar(k-k_+)$ (for example). The Pauli matrices $\sigma_j$ act in the space of the pair of bands that approach each other and touch at the Weyl nodes. The energy spectrum then is $E(p)=v_F|{\bf p}|$ for both nodes. At each node we can associate a chirality, which  measures the relative handedness of the three momenta and the Pauli matrices associated in the Weyl equation. The chirality is $\pm 1$ for the Hamiltonians $H_\pm$. This is a general property of Weyl Fermions realized in band structures - their net chirality must cancel. A simple physical proof of this Fermion doubling theorem is pointed out below.  In a clean system, where crystal momentum is well defined, one can focus at one or the other  node and hence effectively realize the Weyl equation. Note, we have assumed that the bands are individually non-degenerate. This requires that either the time reversal symmetry, or the inversion symmetry (parity), is broken. In order to realize the minimal case of just a pair of opposite chirality Weyl nodes, time reversal symmetry must be broken\cite{Wan}. In practice this is achieved by magnetic order in the crystal. Alternately, one may consider systems with broken inversion symmetry\cite{Murakami}, where a minimum of four Weyl nodes are present.

It is useful to describe a toy lattice model where the above dispersion is simply realized\cite{Ran,David}. Consider electrons hopping on a cubic lattice, where on every site the electron can be spin up or down. Now, assume a spin-orbit type hopping in the $y$ and $z$ directions which proceeds by flipping spin, while along the $x$ direction the sign of hopping depends on the spin projection. The coresponding Hamiltonian is
\begin{equation}
H(k) = \frac{\hbar v_F}{a}\left([\cos (k_xa)+m\left(2-\cos(k_ya) -\cos(k_za)\right)]\sigma_1 + \sin(k_ya) \sigma_2+\sin(k_za) \sigma_3 \right).
\label{eq:WeylModel}
\end{equation}
This Hamiltonian has Weyl nodes located at $(\pm \pi/2a,\,0,\, 0)$. Linearizing about these points yields the Weyl equation \ref{eq:isotropicWeyl}.
Note, one can add an arbitrary (Hermitian) term to this Hamiltonian, which will cause the nodes to shift but cannot remove them for small perturbations.
For example, a Zeeman field $\Delta H=-\left(\hbar v_F/a\right) h_Z \sigma_1$ shifts the nodes to $(k_\pm,\,0,\, 0)$ where $k_\pm a = \pm \cos^{-1} \,h_Z$. Essentially, this stability to perturbations arises from the fact that there is no `fourth' Pauli matrix available to gap out the node. Only when the field $h_Z$ is large enough $|h_Z| \geq 1$ to move the Weyl point up against each other do they annihilate, leading to a fully gapped insulator.

\subsection{Topological Properties}
The stability of the Weyl nodes is tied to a topological protection inherent to this band structure. Away from the band touching points, there is a clear demarcation between filled and empty bands. Consider the state obtained at a particular crystal momentum by filling the negative energy states (below the chemical potential). By studying how this state evolves on varying the crystal momenta one can extract a  Berry phase, from which a Berry flux ${\mathcal B}(k)=\nabla_k\times {\mathcal A}(k)$ can be defined. The Weyl nodes are sources, or monopoles,  of Berry flux - thus $\nabla \cdot {\mathcal B}(k)=\pm \delta^3(k-k_\pm)$, hence their stability. They can only disappear by annihilating a monopole of the opposite charge - which is a Weyl node of opposite chirality\cite{BalentsViewpoint}.

\begin{figure}[t]
\includegraphics[width=.8\textwidth]{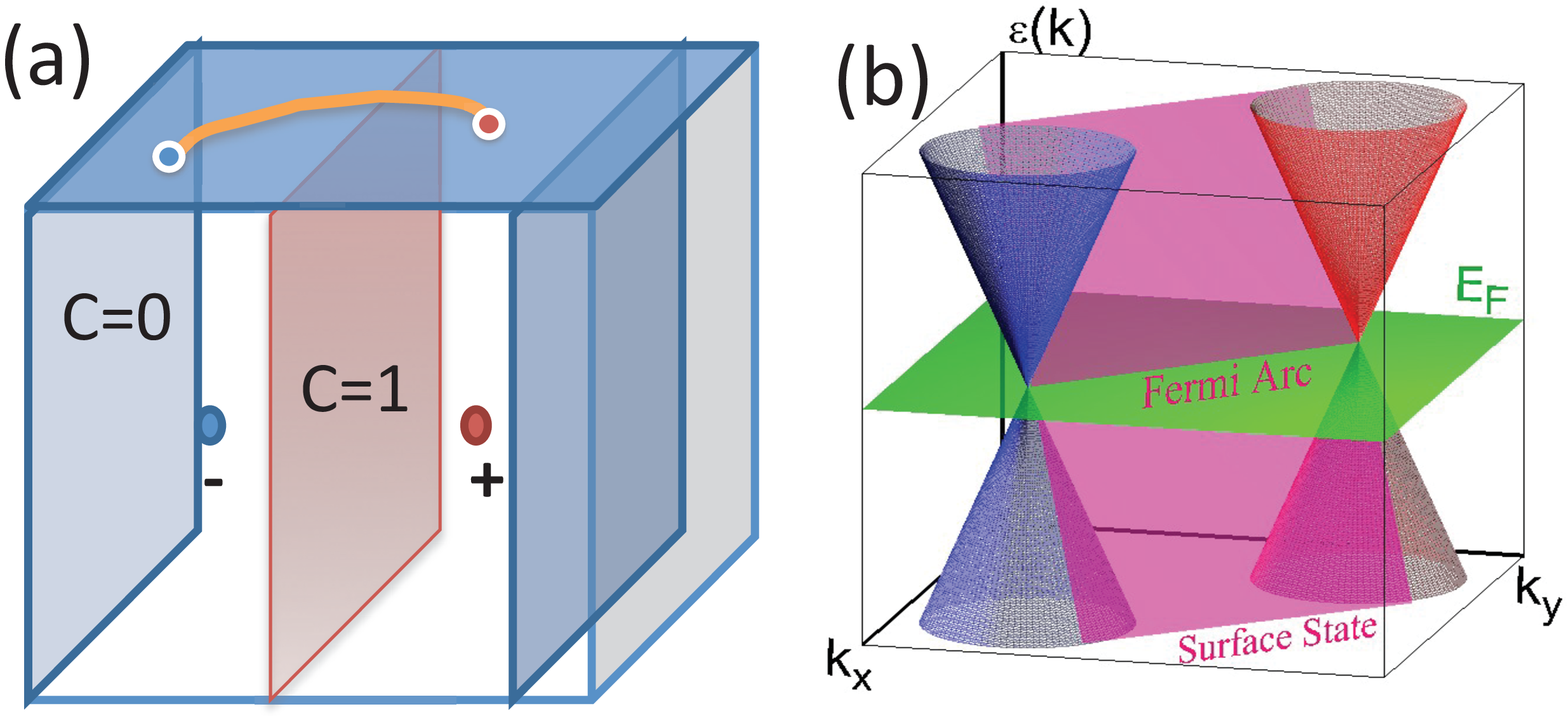}
\caption{\label{fig:Weyl} Weyl semimetal  a)The Fermi arc surface
states of a Weyl semimetal. b) The bulk dispersion (red and blue cones) resolve the paradoxes associated with having a Fermi arc states  (shown in pink) \cite{Wan}. Therefore Fermi arcs are allowed as surface states of a topological semi-metal, but are not possible in free Fermion band structures in 2D.}
\end{figure}

This band topology of Weyl semimetals has two direct physical consequences. The first is an unusual type of surface state, unique to Weyl semimetals - called {\em Fermi arcs} \cite{Wan}. Consider a 3D slab of Weyl semimetal with a surface in the x-y plane. Translation invariance along these directions allow us to label single electron states by crystal momenta in this plane. Let us assume we have a single pair of Weyl nodes in the bulk as in the model in Eq. \ref{eq:WeylModel}. At this same energy we can ask what are the surface states in the system. Surface states are well defined at this energy at all momenta away from the Weyl nodes, because there are no bulk excitations with the same energy and momenta.  It is easily seen that surface states should form a Fermi arc. The arc terminates at the crystal momenta corresponding to the bulk Weyl nodes (see Figure \ref{fig:Weyl}). This result follows from the fact that Weyl nodes are monopoles of Berry flux. Therefore, the 2D Brillouin zones that lie between the pair of Weyl nodes will have a different Chern number than the planes outside (see Figure \ref{fig:Weyl}). These planes may be interpreted as 2D Quantum Hall states associated with a chiral edge state which is guaranteed to cross the chemical potential. The locus of these crossings gives the Fermi Arc surface state.

If one considers both top and bottom surfaces of a Weyl semimetal one should recover a closed Fermi surface as one would expect for a 2D system. Indeed the two Fermi arc states on opposite surfaces, taken together, form a closed 2D Fermi surface. Thus a thin slab of semimetal may be viewed as a 2D system with a closed Fermi surface. As the thickness is increased, two halves of this Fermi surface are spatially separated to opposite sides of the sample. Probing these surface states in surface sensitive probes such as ARPES and STM should provide smoking gun evidence for this unusual phase of matter.

A second physical consequence of the topology of Weyl nodes is their response to an applied electric and magnetic field. As discussed in Section \ref{sec:external perturbations}, a single Weyl node possesses a Chiral Anomaly: the net number of charged particles would not be conserved if a single Weyl node was present\cite{Adler1969}\cite{BellJackiw1969}. Rather the continuity equation is modified
$$
\frac{\partial n}{\partial t}+\nabla\cdot J =\pm \frac{1}{4\pi^2}\frac{e^2}{\hbar^2c} \bE\cdot \bB,
$$
where the sign is determined by the chirality of the Weyl node. Thus charge conservation provides a rationale for why Weyl nodes always must occur in a band structure with  zero net chirality.  Although the net charge is then conserved, the chiral anomaly does lead to an interesting effect. Consider for example the case of a pair of nodes with opposite chirality as in Equation \ref{eq:WeylModel}. Then the difference in density between excitations near the two nodes (the valley polarization) is governed by
\begin{equation}
\frac{d(n_+-n_-)}{dt} = \frac{1}{2\pi^2}\frac{e^2}{\hbar^2c} \bE\cdot \bB
\end{equation}
thus, applying parallel electric and magnetic fields can be used to control the valley polarization - which will lead to new transport phenomena and possibly even applications for Weyl semimetals. There are close connections between this phenomena and the chiral hydrodynamics recently described in the high energy literature\cite{Son1}.  A related physical effect is a giant anomalous Hall effect expected for the case of a pair of Weyl nodes which is proportional to the separation between the Weyl nodes in momentum space. Thus $\sigma_{yz}=\frac{e^2}{2\pi h}(k_+-k_-)$. If combined with an independent measurement of the momentum separation $(k_+-k_-)$ between Weyl nodes, obtained for example via ARPES, leads to a quantized ratio. In Weyl semimetals with higher symmetry, such as cubic symmetry, the anomalous Hall conductance vanishes. However, under a uniaxial strain that lowers symmetry, a large anomalous Hall effect is expected\cite{Ran}.

We note that the two topological properties mentioned above required that the Weyl nodes be separated in crystal momentum. In the presence of breaking of crystalline translation symmetry, such a distinction may be lost, which would obstruct defining a sharp physical property that reflects the underlying topology. Thus it appears that while semimetals like the Weyl semimetal may be topological states, the topology associated with them is sharply defined in the presence of translation symmetry, in contrast to insulating topological phases which do not require any such assumption. However, in practice disorder is rarely strong enough to completely destroy well separated nodal points,  as evidenced in the example of graphene. Thus realistic systems should display the novel features we mentioned above.
\subsection{Physical Realizations}
Despite being a very natural band structure, currently there are no
clearly established materials with Weyl nodes near the chemical
potential, although several promising candidates exist. It has been
proposed that members of the family of material $A_2 Ir_2 O_7$
(pyrochlore iridates), where $A=Y$ or a rare earth such as $A=Eu,\,
Nd,\,Sm$ may be in or proximate to the Weyl semimetal
phase\cite{Wan}. This is currently an active area of experimental
work \cite{Maeno}\cite{Taira}\cite{Zhao}\cite{Tafti}. Spinels based
on osmium\cite{Wan2} and HgCr$_2$Se$_4$\cite{Xu} have also been
proposed as candidates. Another route has been to try to engineer
Weyl semimetals using heterostructures of topological
insulators\cite{Burkov,Halasz}. Interestingly, a proposal to realize
Weyl points in a photonics band structure has recently
appeared\cite{Lu}. A general symmetry analysis of crystal structures
that may host Weyl semimetals appeared in \cite{Manes}. Further
details on this topic may be found in the longer review
\cite{TurnerAV}.

\section{Summary}
We reviewed general conditions under which one may expect gapless
Dirac points to occur in solids. Their appearance may be a
consequence of band-structure effects, of symmetry breaking due to
many-body effects such as superconductivity or as a surface state of
a bulk topological phase. If a Dirac point exists, additional
fine-tuning of the chemical potential is necessary in order for the Dirac point
to coincide with the Fermi level, unless the Dirac point appears as
a consequence of the condensation of Cooper pairs. Then, the Dirac
point ``rides'' along with the chemical potential.

We also reviewed how the Dirac Fermions respond to externally
applied perturbations and why the response differs in the case of
graphene, topological insulators, Weyl semimetals, and d-wave
superconductors. External potentials cause a redistribution of the
quasiparticle spectral weight: space-dependent electrical potential
tends to transfer the spectral weight from large energies towards
the Dirac point, while the Dirac mass term tends to remove the
states from the vicinity of the Dirac point, pushing them towards
the large energies. Uniform magnetic field redistributes the states
over the energy scale set by the cyclotron frequency. In this
context, the magnetic field induced enhancement of the low
temperature specific heat in the vortex state of d-wave
superconductors is also reviewed.

When weak, finite range electron-electron interactions result in
only finite renormalization of the Dirac particle dispersion,
without leading to any qualitative changes. As the strength of the
interactions increases, a quantum phase transition occurs into an
insulating state. In the case of the half-filled repulsive Hubbard
model on the honeycomb lattice, the transition appears to be into a
Neel anti-ferromagnetic state. Among its attractive features is the
possibility to study the transition either using a quantum Monte
Carlo method {\it without} the Fermion sign problem, or analytically using
the $\epsilon$-expansion around $3+1$ dimensions for the continuum
field theory description with the massless Dirac particles Yukawa
coupled to a self-interacting $O(3)$ bosonic field. Understanding
why an interacting system may undergo a symmetry breaking transition
into a state with massless Dirac Fermions, such as in the cuprate
superconductors, rather than avail of a fully gapped state, remains
a fascinating open problem.

Finally we discussed recent developments of three dimensional Weyl
Fermions, including their robust topological properties in the form
of unusual surface states and magneto-electric responses, and
possible physical realizations.

\section{Acknowledgments}
OV was supported by the NSF CAREER award under
Grant No. DMR-0955561, NSF Cooperative Agreement No. DMR-0654118, and the State of Florida.
AV was supported by ARO MURI grant W911NF-12-0461.

\end{document}